\documentclass[reprint,
 amsmath,amssymb,
 aps,pre,
]{revtex4-2}
\usepackage{csquotes}
\usepackage[caption=false]{subfig}
\usepackage{makecell}
\usepackage{graphicx}
\usepackage{dcolumn}
\usepackage{multirow}
\usepackage{bm}

\begin{document}

\preprint{APS/123-QED}

\title{Site percolation in distorted square and simple cubic lattices with flexible number of neighbors}
\author{Sayantan Mitra}
\author{Ankur Sensharma}
\email{itsankur@ugb.ac.in}
 \affiliation{Department of Physics, University of Gour Banga, Malda - 732103, West Bengal, India.}

\date{\today}

\begin{abstract}
This paper exhibits a Monte Carlo study on site percolation using the Newmann-Ziff algorithm in distorted square and simple cubic lattices where each site is allowed to be directly linked with any other site if the euclidean separation between the pair is at most a certain distance $d$, called the connection threshold. Distorted lattices are formed from regular lattices by a random but controlled dislocation of the sites with the help of a parameter $\alpha$, called the distortion parameter. The distinctive feature of this study is the relaxation of the restriction of forming bonds with only the nearest neighbors. Owing to this flexibility and the intricate interplay between the two parameters $\alpha$ and $d$, the site percolation threshold may either increase or decrease with distortion. The dependence of the percolation threshold on the average degree of a site has been explored to show that the obtained results are consistent with those on percolation in regular lattices with extended neighborhood and continuum percolation.  
\end{abstract}
\maketitle
\section{Introduction}
The basic mathematical construction of percolation was introduced by Broadbent and Hammersley in 1957 \cite{Broadbent}. Researchers in various fields have since utilized the basic theory with necessary improvisations to apply it to many natural processes like liquid flow through porous media \cite{Hunt}, forest-fire spreading \cite{Albano1,Guisoni}, epidemic outbreaks \cite{Moore, Ziff2, Miller} and many more \cite{Saberi1,Stauffer,Saberi3}. Percolation has been used by physicists to investigate critical behavior in phase transitions such as the metal-insulator transition \cite{Ball}, the diamagnetic-ferromagnetic transition \cite{Dotsenko} and others.

In the site percolation model, a random occupation mechanism for sites is executed along with a cluster building process through the occupied neighboring sites. As the occupation probability $p$ increases, smaller clusters join together to make larger ones. At the percolation threshold $p=p_c$, the largest cluster spans the lattice. This marks the occurrence of an infinite cluster and signals a phase transition. Another basic variant is the bond percolation model \cite{Broadbent,Wang,Manna}, where the bonds are occupied randomly instead of sites and the spanning cluster is formed by the occupied bonds. Apart from these two basic variants, numerous other models like site-bond percolation \cite{Tarasevich}, directed percolation \cite{Takeuchi}, bootstrap percolation \cite{Adler}, explosive percolation \cite{Adler,Achlioptas}, correlated percolation \cite{Coniglio}, etc. have been designed and applied in diverse fields. Another important model, that bears most resemblance with ours, is perhaps the continuum percolation \cite{Hall1985, Mertens}. In this variant, the spanning is established by randomly located overlapping objects such as sticks, discs, spheres, squares, etc. In Sec. \ref{pc_z}, we have demonstrated the resemblance of our model with continuum percolation when the object is disc or sphere. 

The site percolation threshold $p_c$ has been calculated precisely for many lattices \cite{Wang,Deng,Ballesteros, Gonzalez, Lorentz, Sommers} and the critical exponents \cite{Liu,Kirkpatrick,Kundu2} have also been determined. It should be noted that only the nearest neighbors are considered in the cluster building mechanism in most of these calculations. Site percolation thresholds for square lattices allowing neighbor bonds with specific higher order neighbors have been studied in the past \cite{Dalton1964, Domb1966, Gouker1983, Diribarne1999, Malarz2005}. Recently, a series of detailed studies has been made in regular lattices with extended neighborhood \cite{Malarz2021, Malarz2022, Xun1,Xun2,Xun3}. 

\begin{figure*}
\subfloat[]{\includegraphics[scale=0.3]{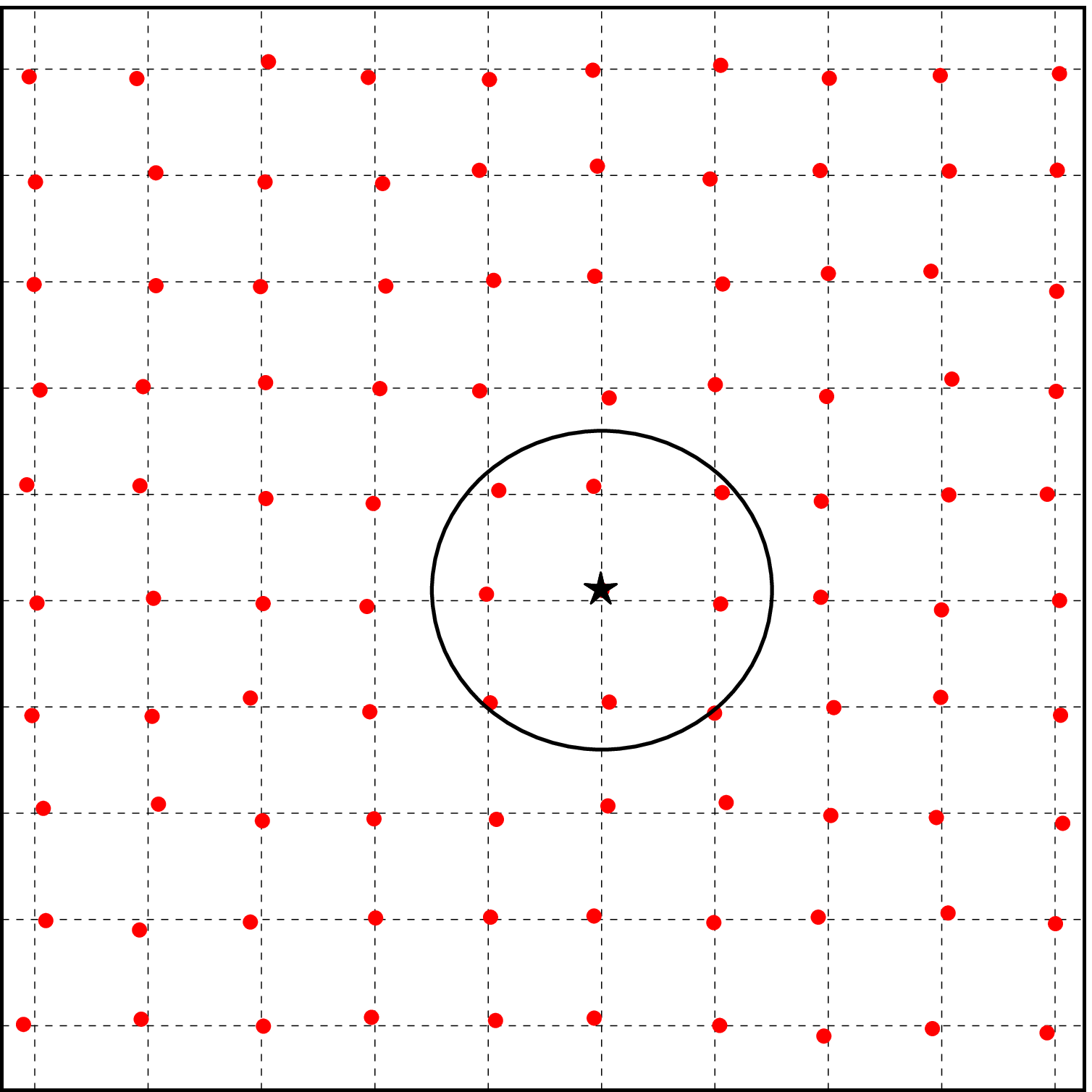}}\hspace{1cm}
\subfloat[]{\includegraphics[scale=0.31]{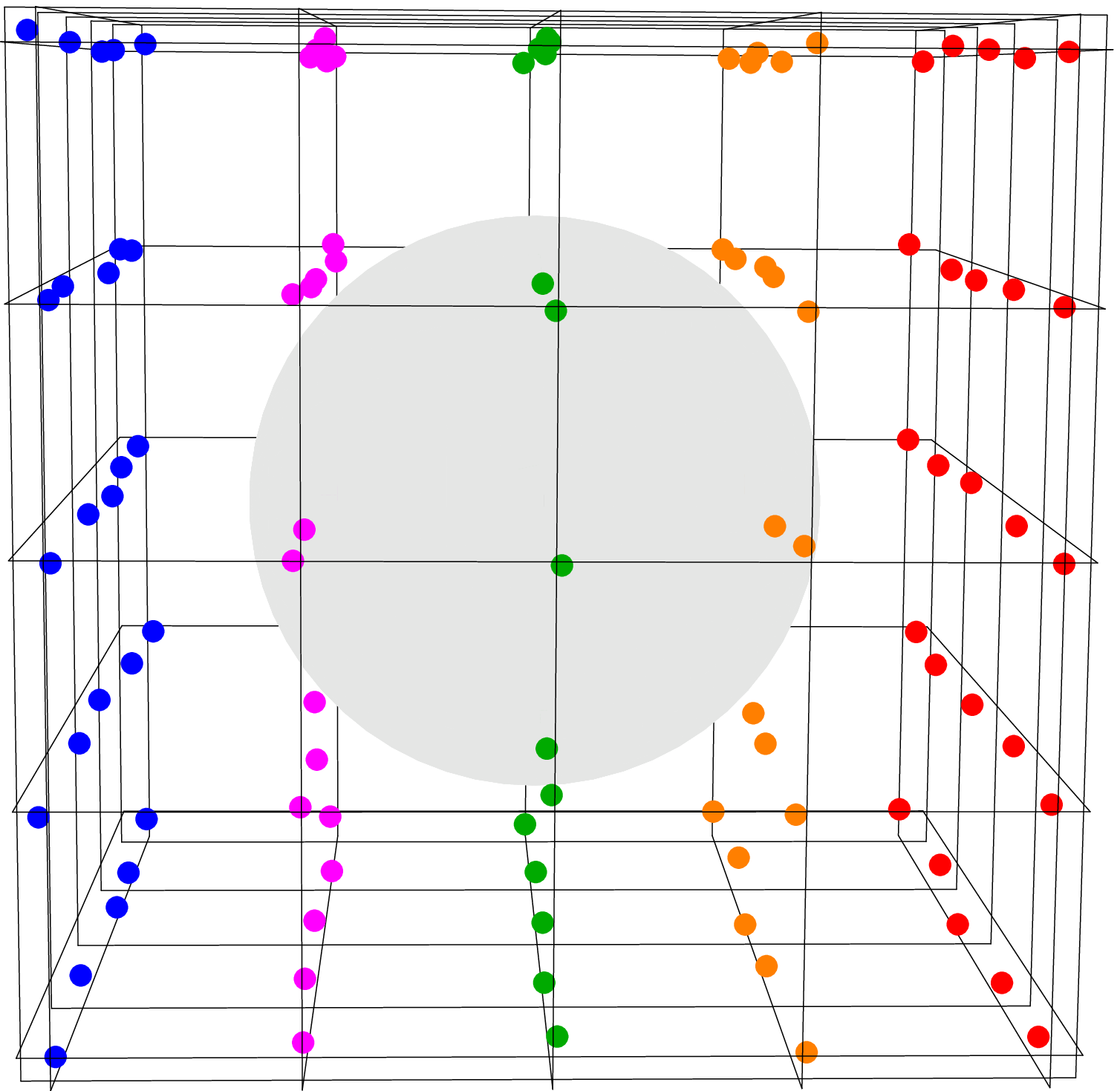}}
\caption{(Color online) (a) A distorted square lattice (DSQL) having $10^2$ sites. The distortion parameter is set at $\alpha=0.1$. The black $\star$ is an occupied site around which a circle is drawn to detect its neighbors. (b) A distorted simple cubic lattice (DSC) having $5^3$ sites with $\alpha=0.1$. The black $\star$ is an occupied site. The neighborhood sphere is displayed to detect its neighbors.}
\label{fig:dist-lattice}
\end{figure*}

Percolation studies in distorted lattices have also been carried out \cite{Sayantan1,Sayantan2,Jang}. These lattices are prepared from the regular lattices by randomly dislocating the sites slightly from their positions. Possible connectivity between the nearest neighbors is subject to the condition $\delta\le d$, where, $\delta$ is the distance between any pair of nearest neighbors and $d$ is a prefixed connection threshold. Since only the nearest neighbors are considered for possible connections, the number of bonds (or, the degree) of a site is $4$ or less for a distorted square lattice and $6$ or less for a distorted simple cubic lattice. 

In this paper, we study site percolation in distorted square and simple cubic lattices where direct bonding with higher order neighbors (next nearest, next next nearest etc.) are allowed if the condition $\delta\le d$ is satisfied. With this relaxation, the degree of a site may also increase from its value for the corresponding regular lattice. The variation of the percolation threshold with $\alpha$ and $d$ exhibits interesting new features. To demonstrate these variations, we have employed the standard Newman-Ziff algorithm \cite{Newman2,Newman1} to calculate the percolation threshold $p_c$ for finite distorted lattices since the percolation thresholds of the corresponding infinite lattices are quite close to these values \cite{Sayantan1, Sayantan2}. 

The rest of the paper is organized as follows: In Sec. \ref{formation} we describe the formation of the distorted lattices and the connection mechanism. We briefly mention the basic simulation process and simulation method in Sec. \ref{simulation}. Sec. \ref{DD} is devoted to demonstrate the resulting distance distribution of the neighboring sites. We present our main results in Secs. \ref{pc_alpha} and \ref{pc_d} showing the variation of the percolation threshold $p_c$ and the average degree $\bar{z}$ with distortion parameter $\alpha$ and the connection threshold $d$. Finally, in Sec. \ref{pc_z}, we establish the consistency of our results with continuum percolation and percolation in regular lattice with extended neighborhood before summarizing.

\section{Model and Method}
\subsection{Distorted lattice formation and the connection criterion}\label{formation}
The distorted square lattice (DSQL) and the distorted simple cubic lattice(DSCL) are formed by randomly dislocating every lattice point from their positions in the regular square lattice (SQL) and simple cubic lattice (SCL) of unit lattice constant respectively. The coordinates of the shifted lattice points may be written as $(x+r_x,y+r_y)$ for DSQL and $(x+r_x,y+r_y,z+r_z)$ for DSCL, where $x, y$, and $z$ are integers between $1$ and $L$, $L$ being the side length of the lattice. The random dislocations in the three directions $r_x$, $r_y$, and $r_z$ range between $-\alpha$ and $+\alpha$, where $\alpha$ is the (tunable) distortion parameter. Thus each dislocated site of a DSQL (DSCL) are found within a square (cube) of length $2\alpha$ centered at its corresponding regular lattice site. See Ref. \cite{Sayantan1,Sayantan2} for details.

Percolation process is established through connections between neighboring sites. The neighbor distances in the regular SQ and SC lattices are just some discrete values. Also, the number of neighbors in every order ({\it e.g.} nearest, next nearest, next next nearest, etc.) are also fixed (see Table I and Table II of \cite{Xun1} for a detailed list). In contrast, the neighbor distances vary in a continuous fashion for the DSQL and DSCL due to distortion. It is, therefore, more meaningful to identify the possible connections by setting a connection threshold $d$, such that, any two sites are connected to each other if and only if their Euclidean distance $\delta$ satisfies $\delta\le d$.

The execution of this connection criterion for the DSQL and DSCL is illustrated in Fig. \ref{fig:dist-lattice}. In a DSQL (DSCL), a circle (sphere) of radius $d$ is drawn around a given site (the $i$-th site, say). This circle (sphere) is called the neighborhood circle (sphere). All the sites within the neighborhood circle (sphere) are then considered to be the neighbors of the $i$-th site (see Fig. \ref{fig:dist-lattice}). 

\begin{figure*}
\subfloat[]{\includegraphics[scale=0.67]{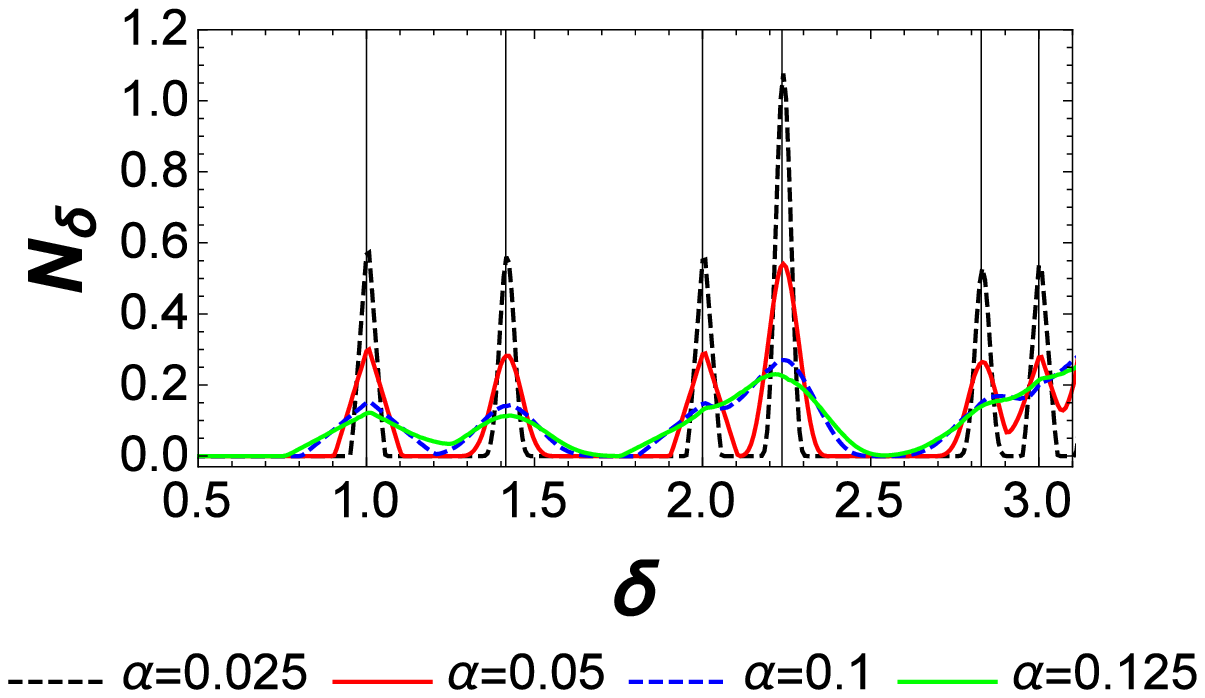}}
\subfloat[]{\includegraphics[scale=0.57]{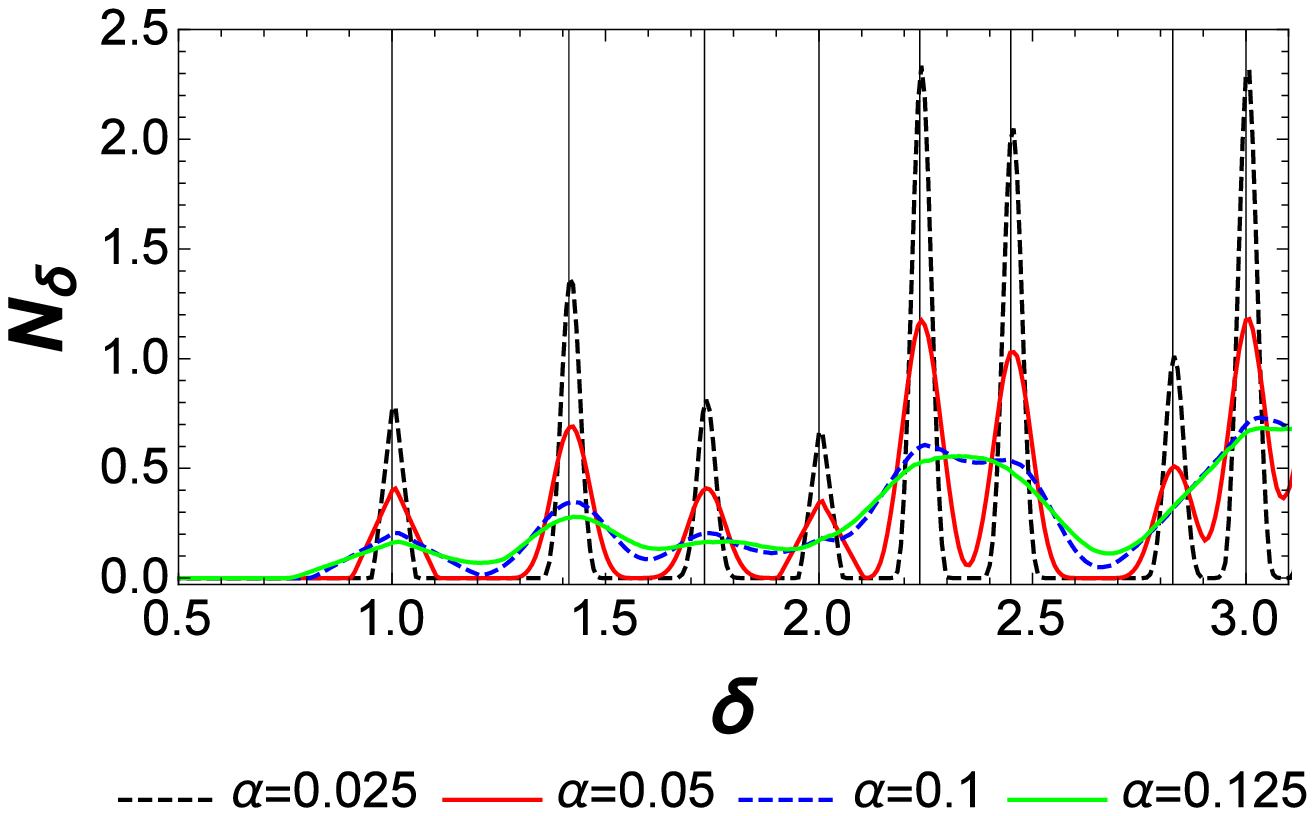}}
\caption{(Color online) Distance distribution of the neighbors for (a) a DSQL with $L=32$ and (b) a DSCL with $L=8$. Different colors in the figures are used for different magnitudes of distortion. The width of bins has been set at $0.008$ for both the histograms.}
\label{fig:delta}
\end{figure*}

\subsection{Simulation method}\label{simulation}
In this section, we briefly outline the basic structure of the Monte Carlo simulations performed to produce the results.
\begin{enumerate}
\item
The DSQL and the DSCL are produced from SQL and SCL by randomly displacing every lattice point as discussed in Sec. \ref{formation}. For each site of a SQ lattice, two independent and identically distributed (i.i.d.) random numbers \footnote{The random number generator used is RAN2()} are generated between $-\alpha$ and $+\alpha$. These are added to the $x$ and the $y$ coordinate of the sites to create the DSQL. The DSCL is created in a similar way using three random numbers for the dislocation of each site.
\item
To obtain the distance distribution of the neighboring sites discussed in Sec. \ref{DD}, the distance $\delta$ between each pair of sites is calculated. These $\delta$ values are then binned into small intervals. The distribution of the sites with distance is then obtained by counting the number of sites in the bins. The counting incorporates an average over $10^3$ independent realizations of both the DSQL and the DSCL.
\item
A percolation study requires the neighbors of each site to be correctly identified. If the distance between a pair of sites meets the condition $\delta \le d$, they are marked as neighbors of each other.
\item
After the detecting the neighbors of each site, we employ the cluster growth mechanism and check the spanning criterion according to the well known Newman-Ziff algorithm \cite{Newman1,Newman2} to determine the percolation thresholds of finite DSQL and DSCL for nearly $500$ different combinations of the distortion parameter $\alpha$ and the connection threshold $d$.
\item
Besides the percolation threshold, we also calculate the average degree $\bar{z}$ of a site from the neighbor section rule, as discussed in step (3). This helps us to analyze how these two quantities are interlinked.
\end{enumerate}

\section{Results and Discussion}   
This section is devoted to the demonstration and discussions of the results obtained from the simulations.

\subsection{Distance distribution of the neighbors}\label{DD}
Before presenting the results of the percolation threshold, it is worthwhile to find out how the distortion affects the average distance distribution of the neighbors.

In Fig. \ref{fig:delta}, we have shown the average number of neighbors $N_\delta$ within a small distance interval $\delta$ and $\delta +d\delta$. For a regular SQ lattice, this plot would show discrete peaks of height $4,4,4,8,4,4,8,...$ at $\delta=1,\sqrt{2},2,\sqrt{5},\sqrt{8},3,\sqrt{10},...$. However, for DSQL, $N_\delta$ varies in a continuous fashion as shown in Fig. \ref{fig:delta}(a). Plots of $N_\delta$ for DSCL with the same set of values of  $\alpha$ are shown in Fig. \ref{fig:delta}(b). The vertical lines are drawn at the distances $\delta_{UD}^{(n)}$ of the neighbors in undistorted SQ and SC lattices. The peaks of the distribution curves for DSQL and DSCL are also located at $\delta = \delta_{UD}^{(n)}$ but the heights of the peaks are lower due to dislocation of sites. Peak-heights can be seen to be approximately proportional to the number of neighbors of various orders. For a fixed number of bins, the distributions are found to obey a normal distribution
\begin{equation}\label{binomialsq}
N(\delta,\alpha) = A_n(\alpha)\ e^{-\sigma(\alpha)\ (\delta_{UD}^{(n)}-\delta)^2},
\end{equation}
where $A_n(\alpha)$ depends on the order $n$ of the neighbors and $\alpha$. $\sigma(\alpha)$ represents the distortion dependent standard deviation, which increases with $\alpha$.

 In Table \ref{tab:delta}, the analytical expressions for maximum $ \delta_M^{(n)}(\alpha)$ and minimum $\delta_m^{(n)}(\alpha)$ distances between the neighbors of various orders $n$ are given as a function of $\alpha$. Here, $n=1,2,...$ means the nearest, next nearest neighbors, and so on. Corresponding neighbor distances $\delta_{UD}^{(n)}$ in undistorted lattices are also given. Since the distance between any two sites may increase or decrease due to distortion, $\delta_m^{(n)}(\alpha)\le \delta_{UD}^{(n)}\le \delta_M^{(n)}(\alpha)$. For example, the fifth nearest neighbor distance of a regular SC lattice is $\sqrt{5}$. But for a DSCL, this may vary between $\sqrt{5-12\alpha+8\alpha^2}\le\delta\le\sqrt{5+12\alpha+12\alpha^2}$.
 
It is clear from Fig. \ref{fig:delta} that the peaks become flatter for higher values of $\alpha$. The successive peaks overlap especially when the difference between the $\delta_{UD}^{(n)}$ values of successive orders of neighbors is small for SQ and SC. For high $\alpha$ values, these overlap portions are increased. The overlap regions indicate that it is possible for a higher-order neighbor to have a low value of $\delta$ (and therefore come closer than a lower order neighbor) and vice versa for distorted lattices.
 
 \begin{figure*}
\subfloat[]{\includegraphics[scale=0.4]{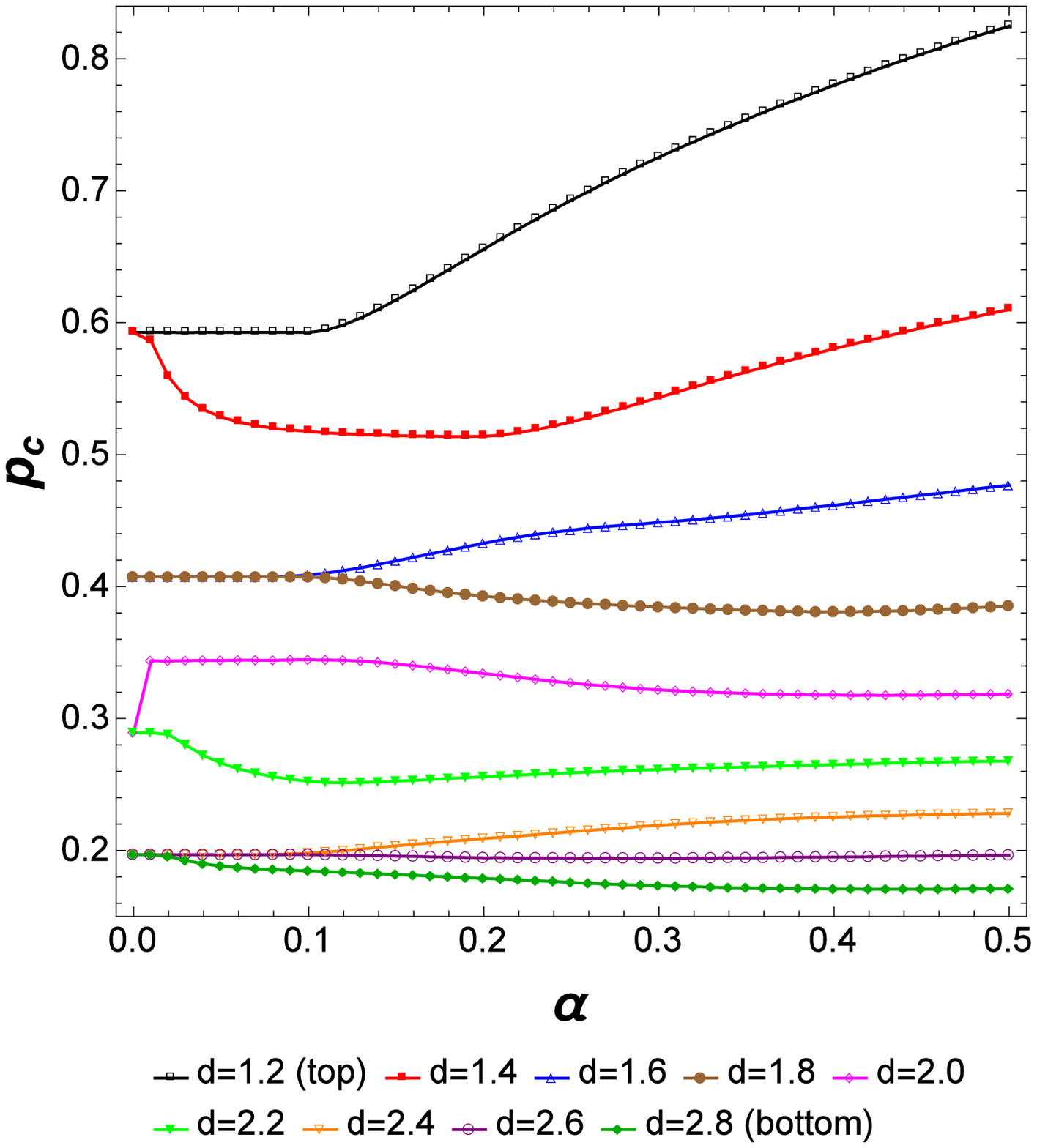}}\hspace{0.3cm}
\subfloat[]{\includegraphics[scale=0.395]{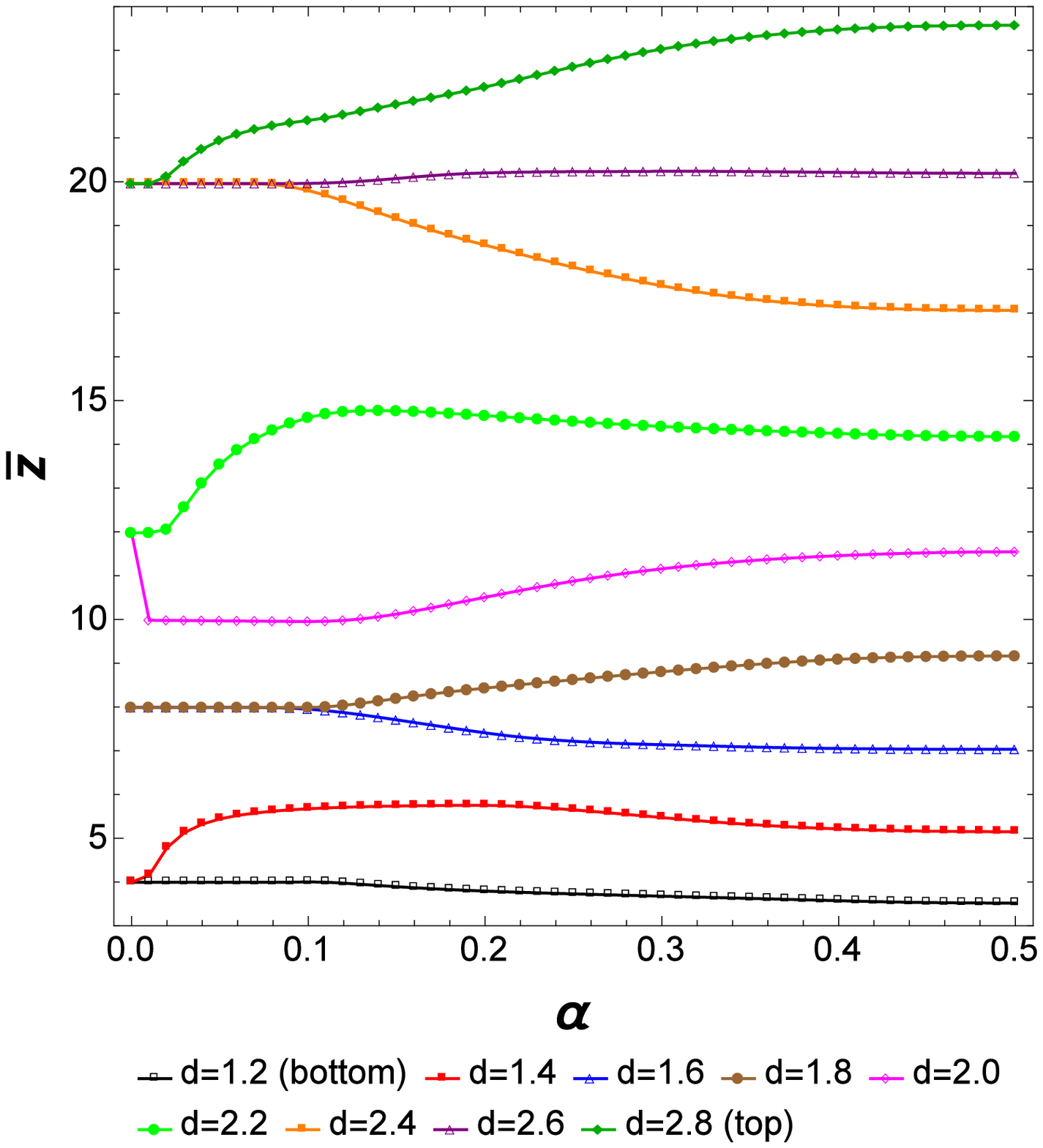}}\\
\subfloat[]{\includegraphics[scale=0.405]{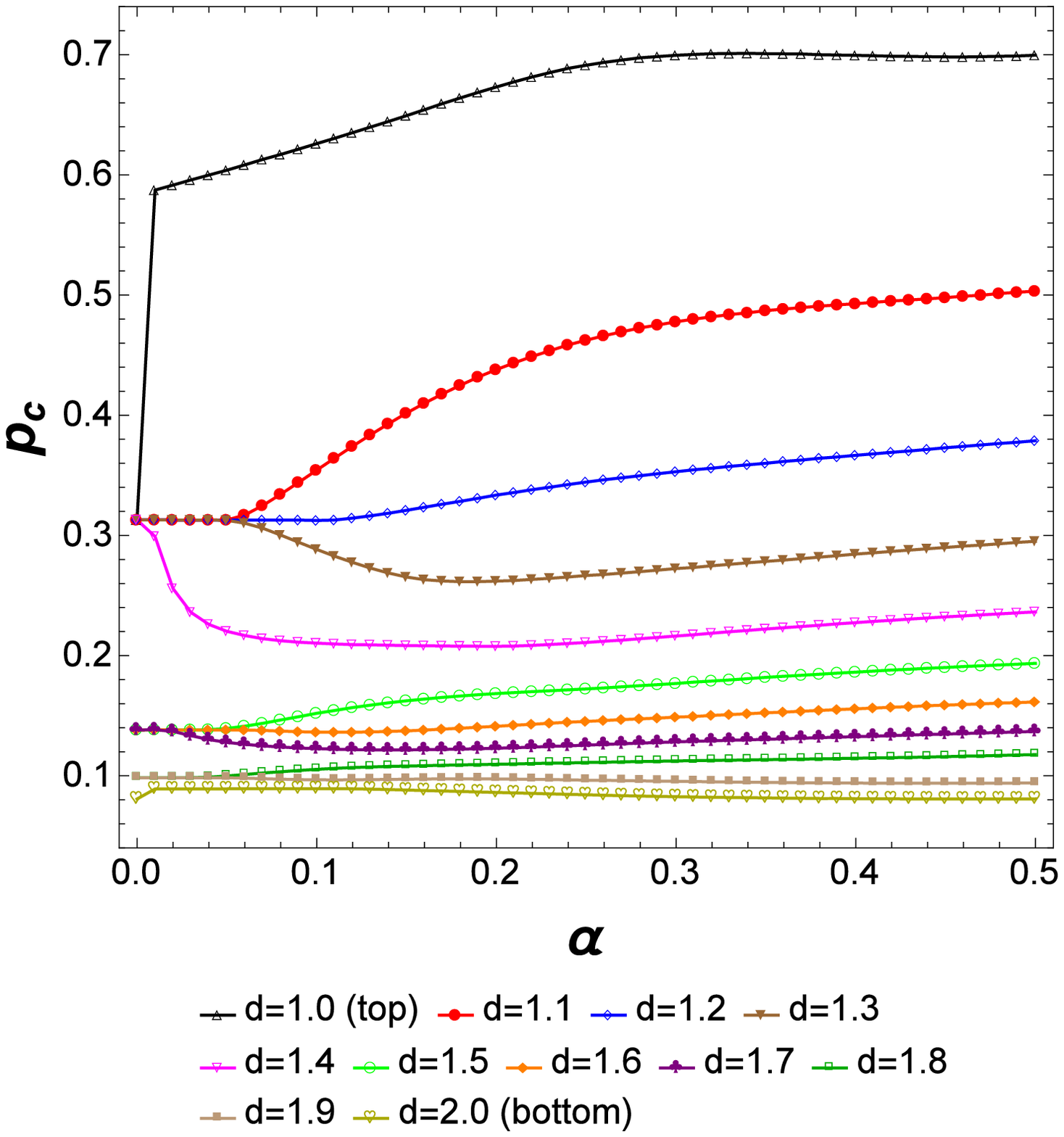}}\hspace{0.3cm}
\subfloat[]{\includegraphics[scale=0.405]{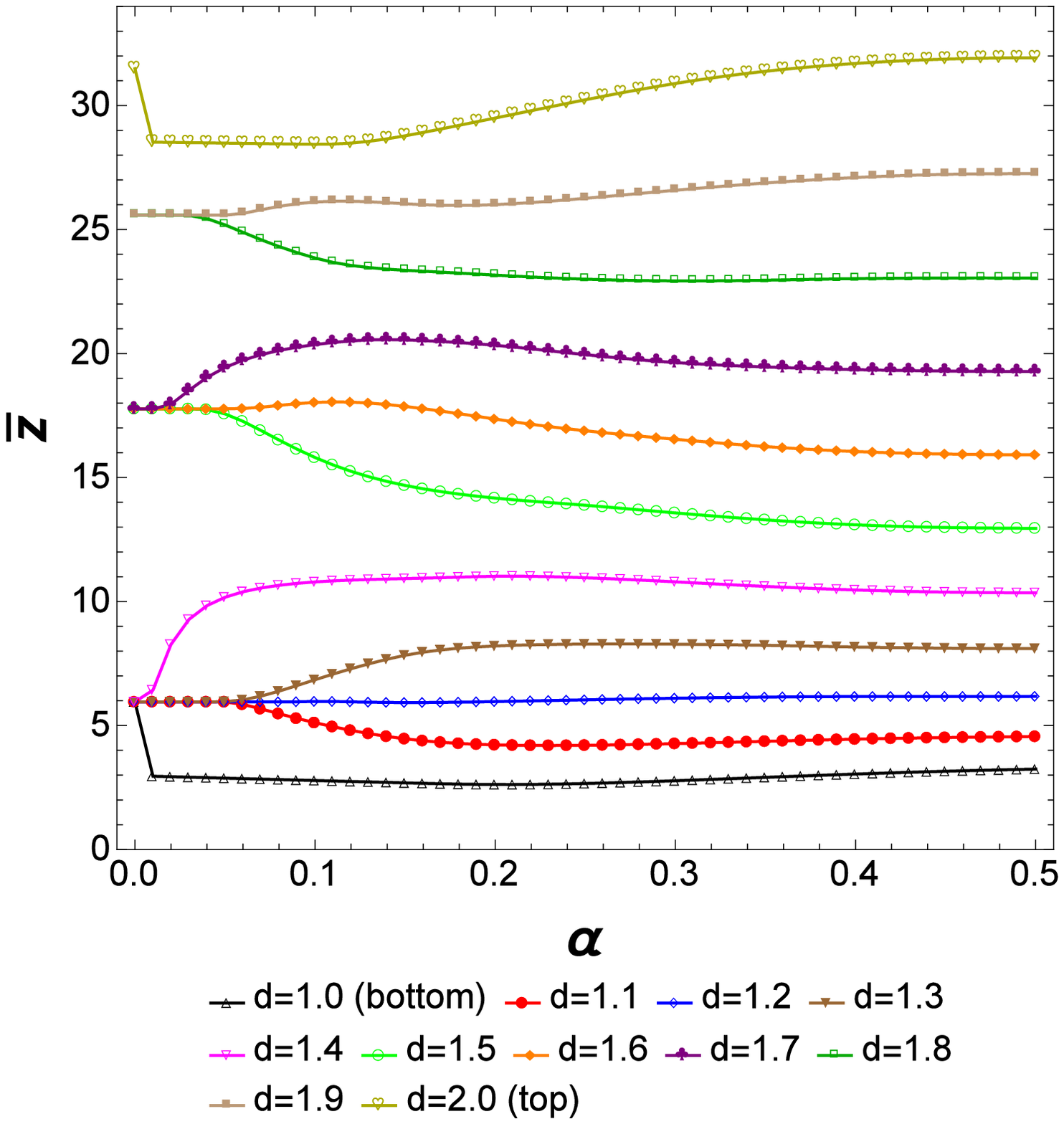}}
\caption{(Color online) $p_c$ and $\bar{z}$ are plotted against $\alpha$ in (a) and (b) respectively for the DSQL of size $L=2^{10}$ using the same set of values of $d$. (c) and (d) show the same for the DSCL of size $L = 2^{7}$. All the data points are obtained by averaging over $10^3$ independent realizations of the lattices. The points are joined by lines to aid viewing.}
\label{fig:alpha-pc-z}
\end{figure*}

\begin{table}[h]
    \begin{center}
    
    \begin{tabular} { | c | c | c | c | c | }
    
    \hline
     Lattice   & Order $n$ & $\delta_{UD}^{(n)}$ & $\delta_M^{(n)}(\alpha)$ & $\delta_m^{(n)}(\alpha)$ \\
    \hline
     DSQL & 1 & 1 & $\sqrt{1+4\alpha+8\alpha^2}$ & $1-2\alpha$ \\
    & 2 & $\sqrt{2}$ & $\sqrt{2} (1+2\alpha)$ & $\sqrt{2} (1-2\alpha)$ \\
    & 3 & 2 & $\sqrt{4+8\alpha+8\alpha^2}$ & $2-2\alpha$ \\
    & 4 & $\sqrt{5}$ & $\sqrt{5+12\alpha+8\alpha^2}$ & $\sqrt{5-12\alpha+8\alpha^2}$ \\
    & 5 & $\sqrt{8}$ & $\sqrt{2} (2+2\alpha)$ & $\sqrt{2} (2-2\alpha)$ \\
    & 6 & 3 & $\sqrt{9+12\alpha+8\alpha^2}$ & $3-2\alpha$ \\
    \hline
     DSCL & 1 & 1 & $\sqrt{1+4\alpha+12\alpha^2}$ & $1-2\alpha$ \\ 
   & 2 & $\sqrt{2}$ & $\sqrt{2+8\alpha+12\alpha^2}$ & $\sqrt{2} (1-2\alpha)$ \\
   & 3 & $\sqrt{3}$ & $\sqrt{3} (1+2\alpha)$ & $\sqrt{3} (1-2\alpha)$ \\
   & 4 & 2 & $\sqrt{4+8\alpha+12\alpha^2}$ & $2-2\alpha$ \\
   & 5 & $\sqrt{5}$ & $\sqrt{5+12\alpha+12\alpha^2}$ & $\sqrt{5-12\alpha+8\alpha^2}$ \\
    \hline
    \end{tabular}
    \end{center}
 \caption{\label{tab:delta} Neighbor distances of first few orders $\delta_{UD}^{(n)}$ for the regular SQ and the SC lattice are shown. The maximum neighbor distances $\delta_M^{(n)}(\alpha)$ and the minimum neighbor distances $\delta_m^{(n)}(\alpha)$ are also tabulated for the DSQL and the DSCL.}
   \end{table}

\begin{figure*}[t]
\subfloat[]{\includegraphics[scale=0.45]{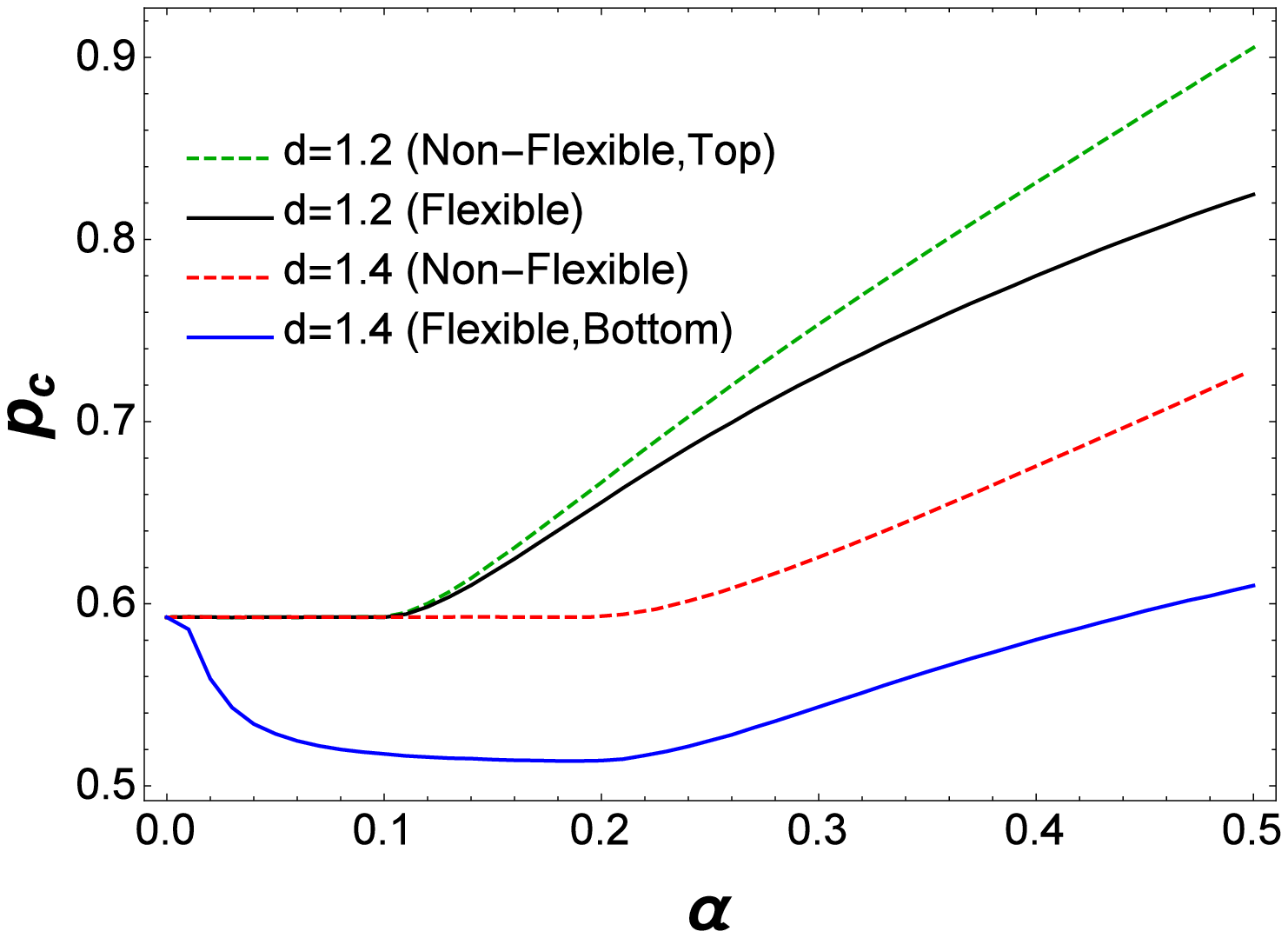}}\hspace{1cm}
\subfloat[]{\includegraphics[scale=0.45]{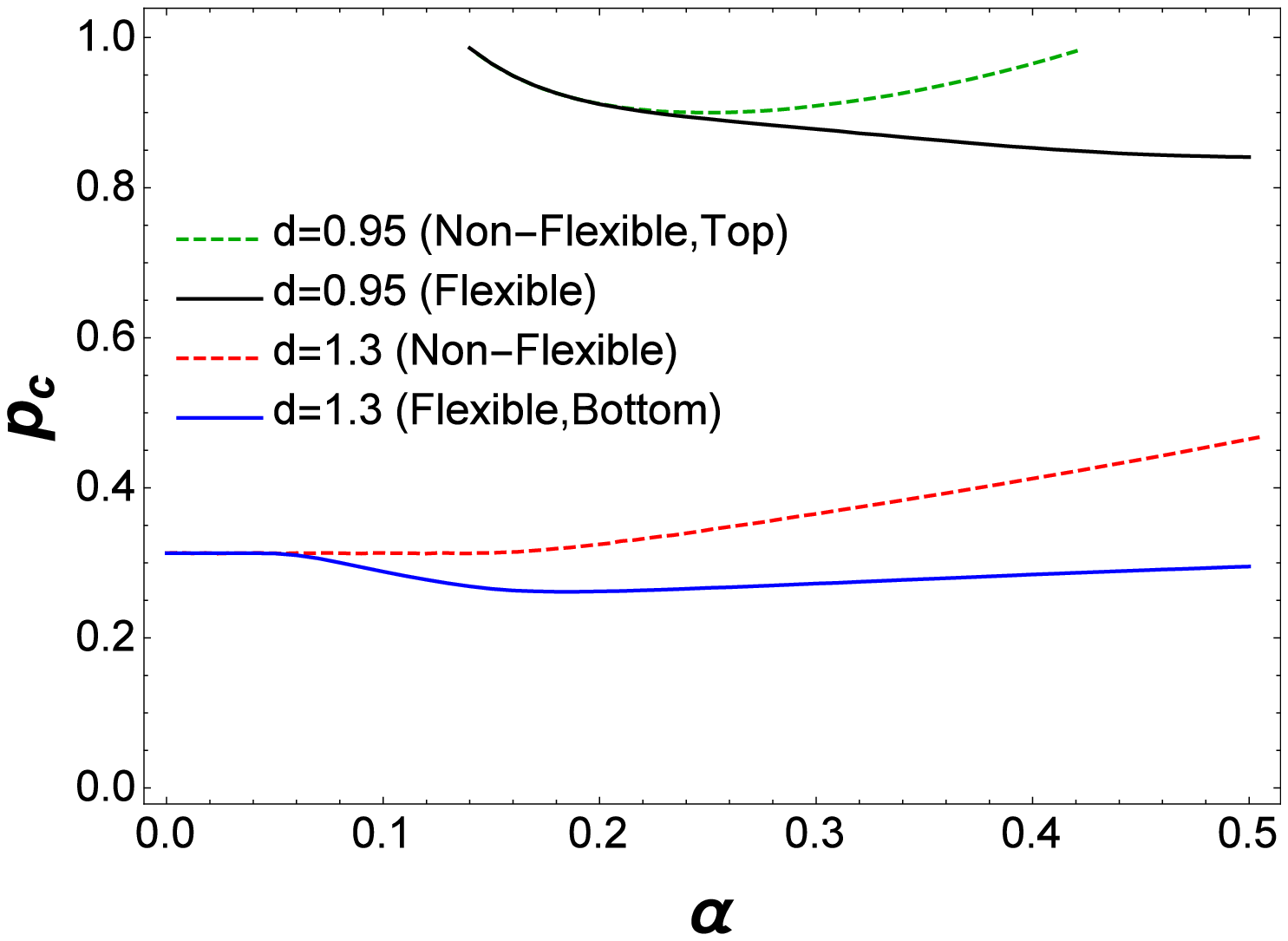}}
\caption{(Color online) Demonstration of how the percolation thresholds of the DSQL (a) and the DSCL (b) change due to relaxation of the restriction. Broken lines are used for non-flexible case, where the bonds are allowed with only the nearest neighbors. Variations of $p_c(\alpha)$ are shown for a pair of values of $d$ for each lattice: $d=1.2$ (top) and $1.4$ for the DSQL (a), and $d=0.95$ (top) and $1.3$ for the DSCL (b). Lattice sizes and number of realizations are the same as in Fig. \ref{fig:alpha-pc-z}}
\label{fig:compare}
\end{figure*}

\subsection{Variation of the percolation threshold}
In an ordinary percolation process, the sites are randomly occupied, and the clusters are developed by the connected occupied neighbors of a lattice. For finite lattices, one looks for the presence of the same cluster in the opposite open-ended boundaries to identify a spanning cluster. The minimum fraction of occupied sites required for spanning is calculated for a large number of independent realizations in order to obtain an average value of the percolation threshold $p_c$ for a finite lattice. In our model, the process is the same except for the neighbor selection rule. The discussion of Sec. \ref{DD} necessitates the distance-dependent rule ($\delta\le d$)to select connecting neighbors of each site. Due to distortion and the connection criterion, the number of neighbors $z$ can in general be different for different sites. Since a spanning path is built through the connection with the neighbors, we have calculated the average number of neighbors, or, the average degree $\bar{z}$ of a site as a function of $\alpha$ and $d$ to demonstrate its impact on $p_c$. 

It should be noted that the possible connections of a site are selected solely based on the distance-dependent connection criterion. The orders of neighborhood have not been taken into consideration. This is the key point of distinction of the present study with the previous ones.

\subsubsection{Variation of $p_c$ with distortion}\label{pc_alpha}
In this section, we demonstrate the impact of distortion on the percolation threshold $p_c$ of DSQL and DSCL for fixed values of the connection threshold $d$. Variations of the average degree $\bar{z}$ with distortion are shown alongside to explain the variation of $p_c$. As expected, $p_c$ and $\bar{z}$ follow reciprocal variation patterns since connections with more neighbors helps spanning and thereby reduces the percolation threshold. We explain the variation patterns of the DSQL only, as similar explanations hold for the DSCL.

\begin{figure*}
\subfloat[]{\includegraphics[scale=0.4]{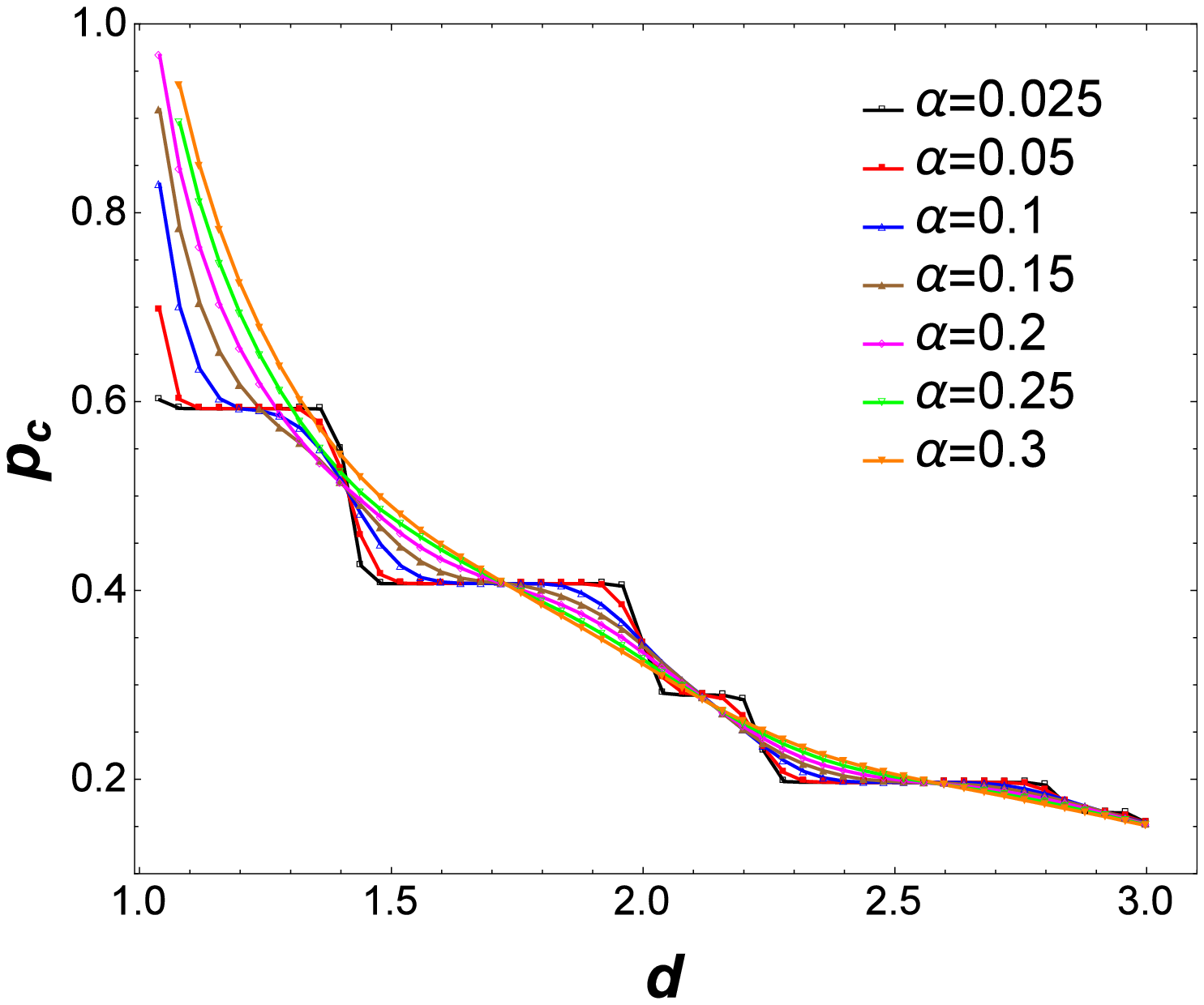}}\hspace{1cm}
\subfloat[]{\includegraphics[scale=0.4]{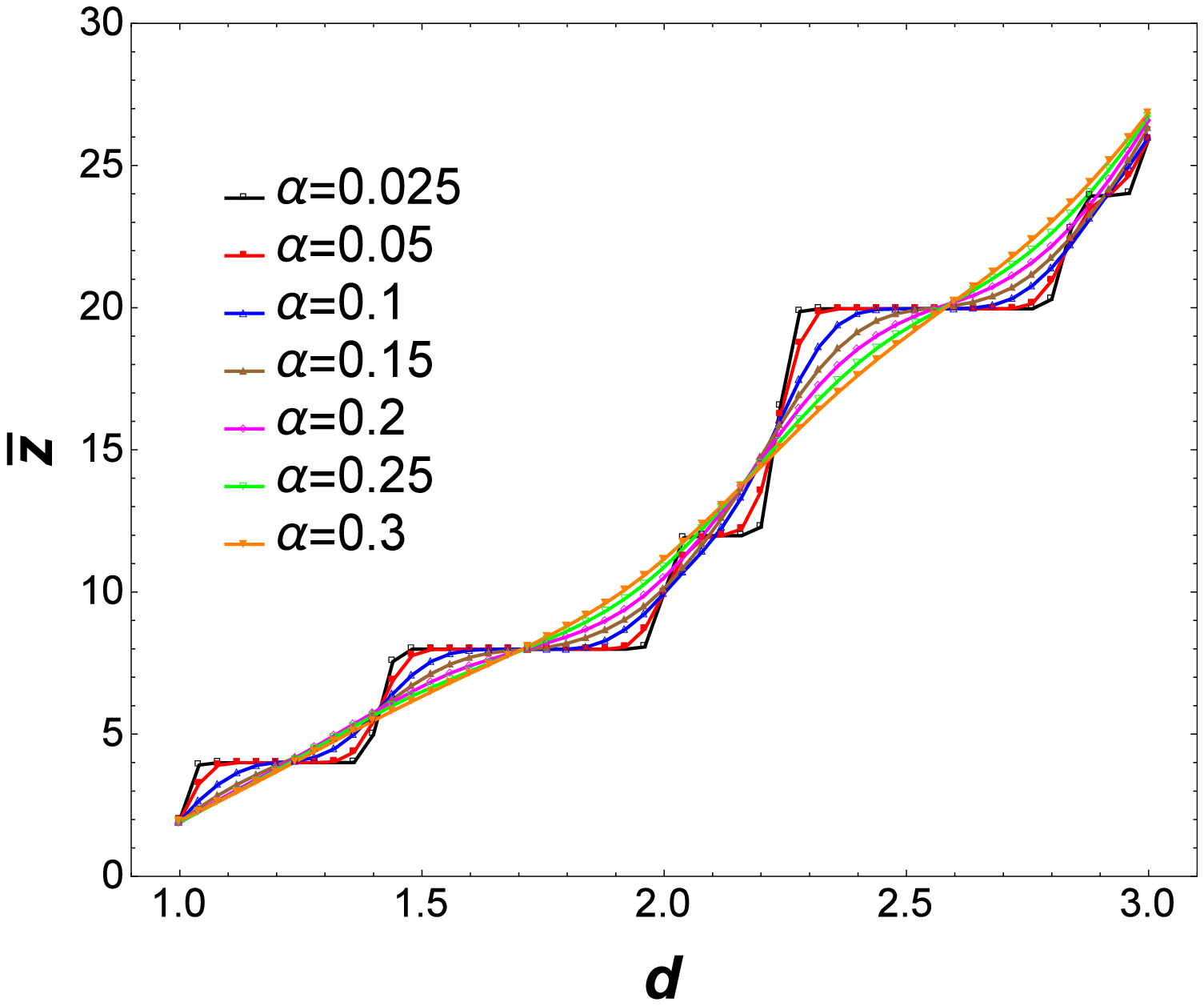}}\\
\subfloat[]{\includegraphics[scale=0.4]{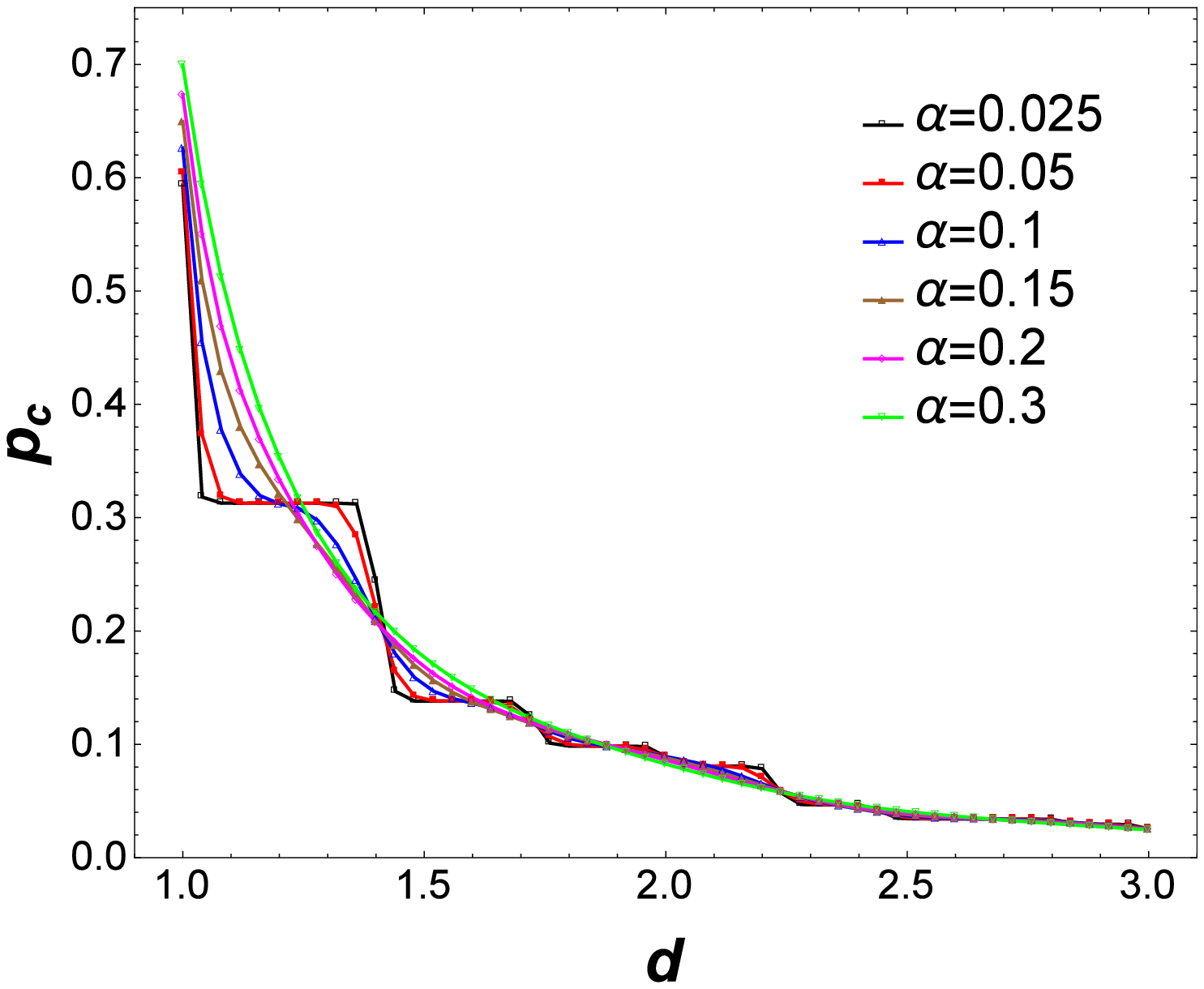}}\hspace{1cm}
\subfloat[]{\includegraphics[scale=0.4]{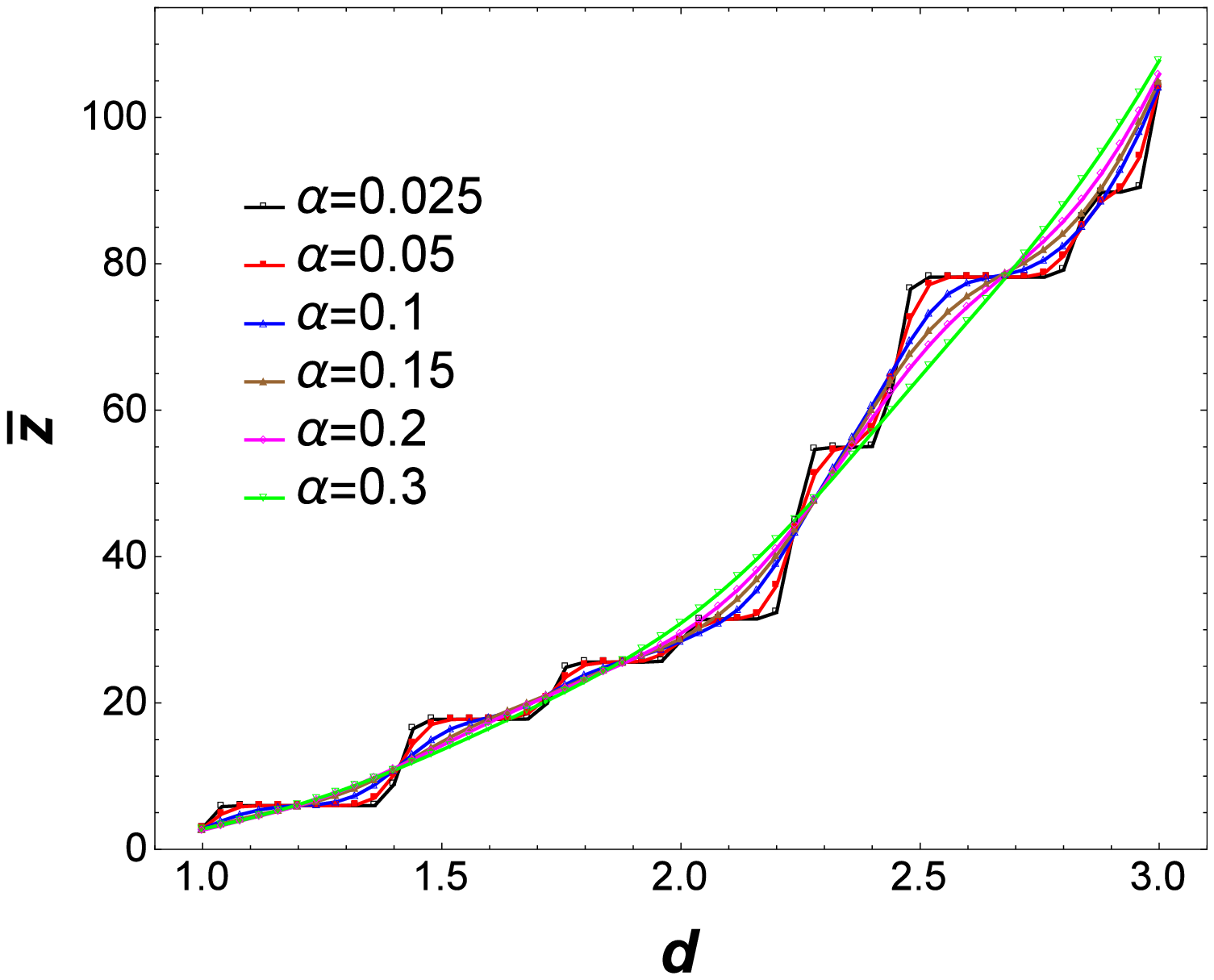}}
\caption{(Color online) $p_c$ and $\bar{z}$ are plotted against $d$ in (a) and (b) respectively for the DSQL of size $L=2^{10}$ using the same set of values of $\alpha$. (c) and (d) show the same for the DSCL of size $L = 2^{7}$. All the data points are obtained by averaging over $10^3$ independent realizations of the lattices. The points are joined by lines to aid viewing. The stair-step pattern is prominent for low distortion and diminishes as distortion is increased.}
\label{fig:nn}
\end{figure*}

Fig. \ref{fig:alpha-pc-z}(a) shows how $p_c$ (here $L=2^{10}$) of DSQL depends on the distortion parameter $\alpha$ when $d$ is held fixed at nine different values. The follwoing observations are made.
\begin{enumerate}
\item At $\alpha=0$, the nine curves of $p_c$ start from four different values, which are very close to the values obtained in \cite{Xun1}. This fact can be understood by observing the $\delta_{UD}^{(n)}$ column of Table \ref{tab:delta} and considering the fact that connection with more neighbors helps to decrease the percolation threshold. For example, the $p_c$ curves corresponding to $d=1.6$ and $d=1.8$ start at the same value because these two values of the connection threshold ensure that only first two neighbors of the regular SQ lattice (with $\delta_{UD}^{(1)}=1$ and $\delta_{UD}^{(2)}=\sqrt{2}$) are encompassed. The curve for $d=1.4$ starts with a higher value since the second nearest neighbors at $\delta_{UD}{(2)}=\sqrt{2}$ are left out. Similarly, the curves for $d=2.0$ and $d=2.2$ start at a lower value because the third order neighbors ($\delta_{UD}^{(3)}=2.0$) are included. 

\item Interestingly, when $d$ is equal to any one of $\delta_{UD}^{(n)}$ values of Table. \ref{tab:delta}, a sudden increase in $p_c$ is observed. This happens because a very small amount of distortion eliminates some of the neighbors resulting in sudden drop in $\bar{z}$ and consequently a sudden jump in $p_c$. For higher values of $\alpha$, some neighbors are included again to reduce $p_c$ slightly. (See the curves of $d=2.0$ in Fig. \ref{fig:alpha-pc-z}(a) and Fig. \ref{fig:alpha-pc-z}(b) for DSQL and those of $d=1.0$ and $2.0$ in Fig. \ref{fig:alpha-pc-z}(c) and \ref{fig:alpha-pc-z}(d) for DSCL).

\item The $p_c(\alpha)$ curves with different values of $d$ follow different patterns although some of them start with the same value at $\alpha=0$. The branching-like patterns of Figs. \ref{fig:alpha-pc-z}(a) and \ref{fig:alpha-pc-z}(c) can be explained in the following way. If the value of $d$ is slightly less than a regular neighbor distance $\delta_{UD}^{(n)}$, a little distortion would increase the average number of neighbors $\bar{z}$ resulting in a reduction of $p_c$. In Fig. \ref{fig:alpha-pc-z}(a), the curves of $d=1.4$ and $d=2.2$ for the DSQL show this feature since these values of $d$ are slightly less than $\delta_{UD}^{(2)}$ and $\delta_{UD}^{(4)}$ respectively. For $d=1.8$, this reduction of $p_c$ starts to occur at $\alpha\approx 0.1$ since $\delta_m^{(3)}(\alpha)$ becomes lower than $d$.

\item In the high $\alpha$ region, the explanations behind rises and falls of $p_c$ become less straightforward due to increased possibility of involvement of higher order neighbors. However, one may analyze case by case remembering two key points. (i) It is clear from Table \ref{tab:delta} that $\delta_M^{(n)}(\alpha)$ is an increasing function of $\alpha$. If $\delta_M^{(n)}(\alpha)<d$, all the neighbors upto order $n$ are included. As $\alpha$ increases, $\delta_M^{(n)}(\alpha)$ approaches $d$. When $\delta_M^{(n)}(\alpha)$ crosses the value of $d$, some of the connections with the neighbors start to break. As a result, $p_c$ increases. (ii) On the contrary, $\delta_m^{(n)}(\alpha)$ is a decreasing function. For a sufficiently large distortion, $\delta_m^{(n)}(\alpha)$ becomes less than $d$, allowing some connections of order $n$ to cause a reduction in $p_c$. 

\item In Fig. \ref{fig:alpha-pc-z}(b) and \ref{fig:alpha-pc-z}(d), $\bar{z}(\alpha)$ is plotted for the same set of values of $d$ used in \ref{fig:alpha-pc-z}(a) and \ref{fig:alpha-pc-z}(c) respectively. As expected, the curves are reversed -- the $d=2.0$ curve goes to the top and the $d=1.0$ curve comes to bottom -- since higher $\bar{z}$ helps spanning. Notice that, the values of $\bar{z}$ at $\alpha=0$ are slightly less than integers due to boundary effect. The nodes near the boundaries of finite lattices have lesser degrees in absence of the periodic boundary conditions.

\item Fig. \ref{fig:compare} shows how the percolation threshold changes due to the relaxation in the connection criterion. Note that, if the bonds are only allowed between nearest neighbors, $p_c$ always increases with distortion when $d$ is greater than the lattice constant ($=1$)\cite{Sayantan1, Sayantan2}. Clearly, this is not the case when the neighbors are selected solely on the basis of euclidean separation. The variation of $p_c(\alpha)$ for $d<1$ in the DSCL is also different. In the DSQL, no spanning cluster can be found for $d\le 1$ even with relaxed connectivity.
\end{enumerate}

\begin{figure*}
\subfloat[]{\includegraphics[scale=0.5]{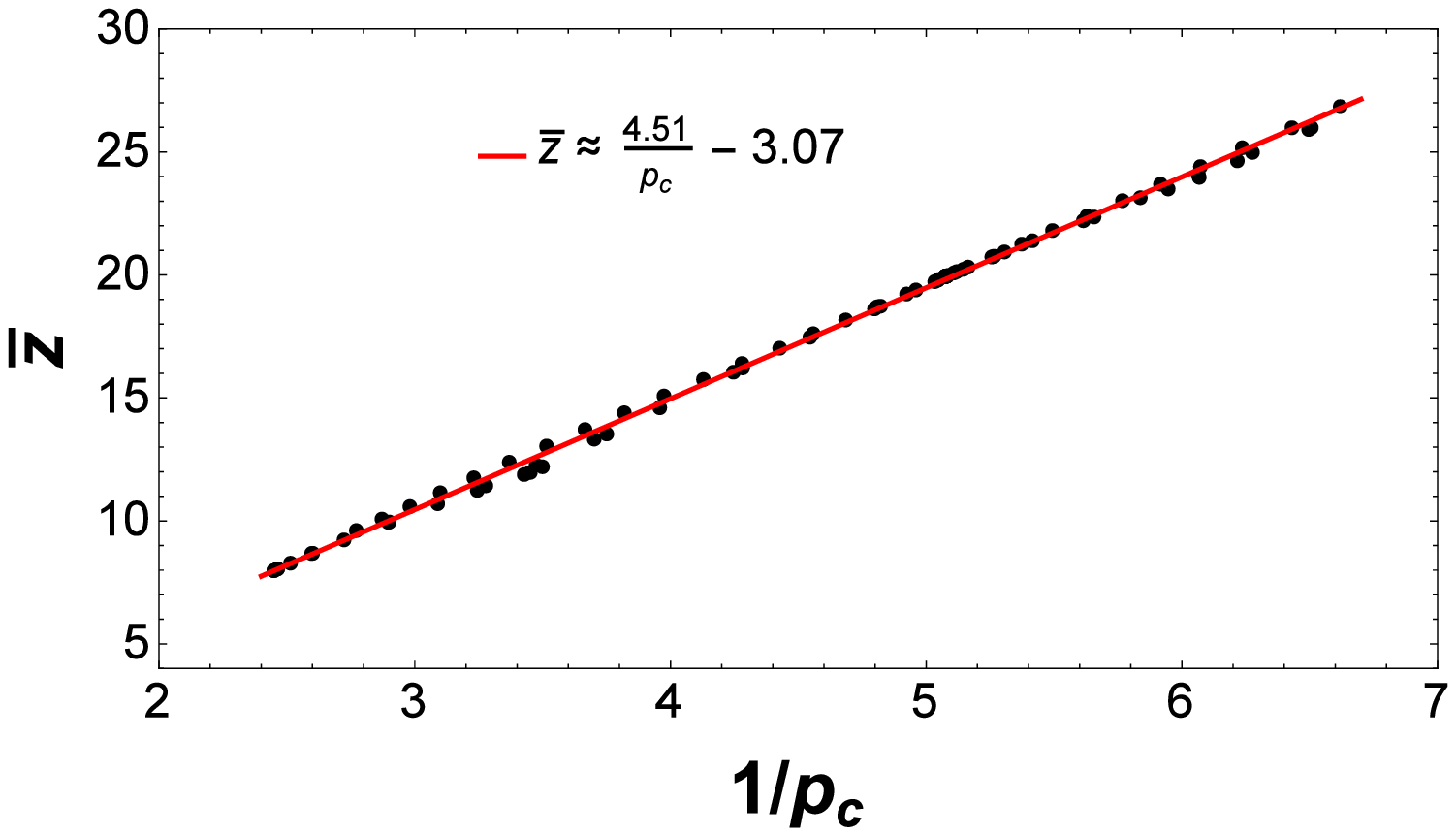}}\hspace{1cm}
\subfloat[]{\includegraphics[scale=0.52]{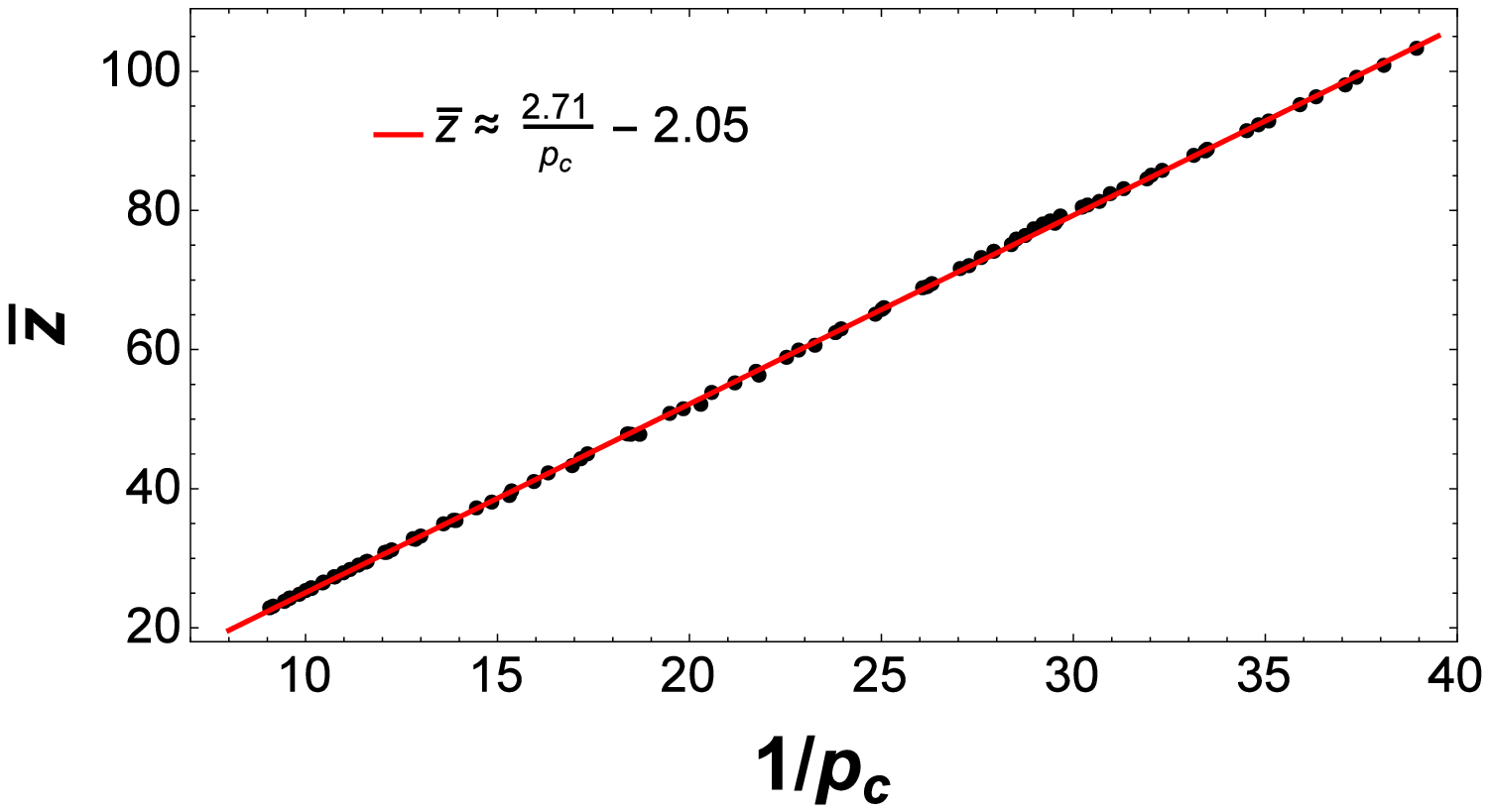}}
\caption{(Color online) $\bar{z}$ is plotted against $1/p_c$ for the DSQL and DSCL in (a) and (b) respectively. Lines are drawn according to the best linear fits written in the figures. Slope of the straight line for DSQL and DSCL are close to the analytical prediction. These data points are collected from the data of Figs. \ref{fig:alpha-pc-z} and \ref{fig:nn}.}
\label{fig:z-pc}
\end{figure*}

\subsubsection{Variation of $p_c$ with connection threshold}\label{pc_d}
The variations of the percolation threshold $p_c$ and the average degree of a site $\bar{z}$ with the connection threshold $d$ for fixed values of $\alpha$  are shown in Fig. \ref{fig:nn}. The patterns can be explained by a logical extension from the $\alpha=0$ case. When distortion is absent, $\bar{z}$ should behave as a piecewise constant function because the neighbors are located at predefined positions. If $\delta_{UD}^{(i)}<d<\delta_{UD}^{(i+1)}$, the neighbors upto order $i$ are taken into consideration. Therefore, $\bar{z}(d)$ increases in steps and the function can be expressed as

\begin{equation}\label{eq:step}
\bar{z}(d)=\sum_i z_i H(d-\delta_{UD}^{(i)}),
\end{equation}
where, $z_i$ is the number of $i$-th order neighbors ($z_1=z_2=z_3=4, z_4=8$ etc. for a regular SQ lattice) and $H(x)$ is the Heaviside step function.

Note that, the plots of $\alpha=0.025$ in Figs. \ref{fig:nn}(b) (DSQL) and \ref{fig:nn}(d) (DSCL) nearly follow Eq. \ref{eq:step} since distortion is small. As $\alpha$ increases, the discrete plateaus gradually vanish and the variations of $\bar{z}$ with $d$ become unwrinkled.

As expected, $p_c$ decreases with $d$ in steps for an undistorted lattice. The variations of $p_c$ and $\bar{z}$ follow inverse patterns of as shown in Figs. \ref{fig:nn}(a) (DSQL) and \ref{fig:nn}(c) (DSCL). 

\subsection{Linkage with other related studies}\label{pc_z}
Having discussed how the percolation threshold and the average degree of a site varies $\alpha$ and $d$, we are now in a position to relate these two quantities for distorted lattices. This has been discussed in a number of studies on continuum percolation \cite{Domb, Mertens, Quintanilla, Tarasevich1, Lorenz1} and on percolation with extended neighborhood \cite{Xun1, Xun2}. 

In the case of continuum percolation, a spanning path is built up through identical overlapping objects. The continuum percolation threshold is defined as : 
\begin{equation}\label{etac}
\eta_c = a_{_D}r^D \left( \frac{N_c}{V}\right)
\end{equation}

where $N_c$ is the critical number of objects required for spanning, $V$ is the volume of the whole system in $D$ dimensions, $r$ represents a length scale of the object (e.g. radius for disc or sphere), and $a_{_D}$ is a constant depending on the shape of the objects. 

The way of selecting the neighbors in DSQL and DSCL suggests that this model should be compared to the continuum percolation model with discs and spheres as objects. A comparison between the connection criterion of the distorted lattices and the overlapping mechanism of the discs or the spheres in continuum percolation reveals that $d=2r$ and $\bar{z} = a_{_D} d^D=a_{_D} r^D 2^D$ . Combining this with Eq. (\ref{etac}) one gets
\begin{equation}\label{z1}
\frac{N_c}{V} = \frac{(2)^D \ \eta_c}{\bar{z}}.
\end{equation}

The connection with the site percolation is usually established \cite{Xun1} by assuming that $N_c$ number of objects of unit volume are required for a spanning path to exist in a volume $V=L^{D}$. This allows one to realize that the role of the percolation threshold $p_c$ and the fraction $\frac{N_c}{V}$ are the same. Eq. (\ref{z1}) then becomes  
\begin{equation}\label{pc}
p_c = \frac{(2)^D \ \eta_c}{\bar{z}} 
\end{equation}

So, for DSQL, one can write  
\begin{equation}\label{pcDSQL}
\begin{split}
p_c & = \frac{(2)^2 \ \eta_c^{disc}}{\bar{z}},\\
\\
    & = \frac{4.51235}{\bar{z}},
\end{split}
\end{equation}

as $\eta_c^{disc} = 1.128087$ \cite{Xun1,Mertens,Quintanilla,Tarasevich1}. Similarly the relation between $p_c$ and $\bar{z}$ for DSCL is : 
\begin{equation}\label{pcDSCL}
\begin{split}
p_c & = \frac{(2)^3 \ \eta_c^{sphere}}{\bar{z}},\\
\\
    & = \frac{2.7351}{\bar{z}},
\end{split}
\end{equation}

as $\eta_c^{sphere} = 0.34189$.

Accounting for the finite-$z$ effect, $\bar{z}$ can be expressed as : 
\begin{equation}\label{finitez}
\bar{z} = \frac{c}{p_c} - b
\end{equation}

In Fig. \ref{fig:z-pc} $(a)$ and $(b)$, $1/p_c$ are plotted against $\bar{z}$ for DSQL and DSCL. The calculations are done for DSQL of side length $L=1024$ and DSCL of side length $L=128$. Each point in the plots represent an average over $10^3$ independent realizations of DSQL and DSCL. The points are chosen from sets used in Figs. \ref{fig:alpha-pc-z} and \ref{fig:nn} irrespective of the values of $\alpha$ and $d$. The results of the best linear fits with Eq. \ref{finitez} yield $c=4.509(17)$ and $c=2.712(4)$ for the DSQL and DSCL respectively. These values are close to the values of $c = 2^D \eta_c$ predicted by the Eqs. (\ref{pcDSQL}) and (\ref{pcDSCL}). Better precision of these values may be obtained by averaging over a large number of realizations of bigger lattices which is beyond the scope of this work.
\section{Summary}
To summarize, we have characterized site percolation by applying the Newman-Ziff algorithm in distorted square and simple cubic lattices where the number of connected neighbors is not fixed. The distortion is introduced in the system through a tunable parameter $\alpha$. The neighbors of a site are selected by the connection rule $\delta \leq d$, where, $\delta$ is the distance between any two sites and $d$ is the connection threshold. In the previous studies on distorted lattices \cite{Sayantan1, Sayantan2}, this connection rule was applied only to the nearest neighbors. Thus the possible number of bonds from a site could be $4$ or less for the SQ lattice, and $6$ or less for the SC lattice. In this study, this number can increase, decrease, or remain the same. We have shown that this flexibility gives rise to an approximately normal distribution of neighbor distances around a regular lattice point yielding interesting new features in the variation of the percolation threshold with $\alpha$ and $d$. Comparison with earlier studies in distroted lattices shows significant change in $p_c$ even for low distortion. Recent percolation studies with extended neighborhood in undistorted lattices \cite{Xun1, Xun2} did not require any flexibility in the number of neighbors since the regular lattice points have predefined locations.  Our results are consistent with those obtained in Ref. \cite{Xun1} in the zero distortion limit. The mapping with the continuum percolation model has been explicitly demonstrated through the dependence of $1/p_c$ on average degree $\bar{z}$. 

\section{Acknowledgment}
We thank Dipa Saha, Bishnu Bhowmik, and Abdur Rashid Miah for useful discussion. The computation facilities availed at the Department of Physics, University of Gour Banga, Malda are gratefully acknowledged. 

\begin{thebibliography}{49}%
\makeatletter
\providecommand \@ifxundefined [1]{%
 \@ifx{#1\undefined}
}%
\providecommand \@ifnum [1]{%
 \ifnum #1\expandafter \@firstoftwo
 \else \expandafter \@secondoftwo
 \fi
}%
\providecommand \@ifx [1]{%
 \ifx #1\expandafter \@firstoftwo
 \else \expandafter \@secondoftwo
 \fi
}%
\providecommand \natexlab [1]{#1}%
\providecommand \enquote  [1]{``#1''}%
\providecommand \bibnamefont  [1]{#1}%
\providecommand \bibfnamefont [1]{#1}%
\providecommand \citenamefont [1]{#1}%
\providecommand \href@noop [0]{\@secondoftwo}%
\providecommand \href [0]{\begingroup \@sanitize@url \@href}%
\providecommand \@href[1]{\@@startlink{#1}\@@href}%
\providecommand \@@href[1]{\endgroup#1\@@endlink}%
\providecommand \@sanitize@url [0]{\catcode `\\12\catcode `\$12\catcode
  `\&12\catcode `\#12\catcode `\^12\catcode `\_12\catcode `\%12\relax}%
\providecommand \@@startlink[1]{}%
\providecommand \@@endlink[0]{}%
\providecommand \url  [0]{\begingroup\@sanitize@url \@url }%
\providecommand \@url [1]{\endgroup\@href {#1}{\urlprefix }}%
\providecommand \urlprefix  [0]{URL }%
\providecommand \Eprint [0]{\href }%
\providecommand \doibase [0]{https://doi.org/}%
\providecommand \selectlanguage [0]{\@gobble}%
\providecommand \bibinfo  [0]{\@secondoftwo}%
\providecommand \bibfield  [0]{\@secondoftwo}%
\providecommand \translation [1]{[#1]}%
\providecommand \BibitemOpen [0]{}%
\providecommand \bibitemStop [0]{}%
\providecommand \bibitemNoStop [0]{.\EOS\space}%
\providecommand \EOS [0]{\spacefactor3000\relax}%
\providecommand \BibitemShut  [1]{\csname bibitem#1\endcsname}%
\let\auto@bib@innerbib\@empty
\bibitem [{\citenamefont {Broadbent}\ and\ \citenamefont
  {Hammersley}(1957)}]{Broadbent}%
  \BibitemOpen
  \bibfield  {author} {\bibinfo {author} {\bibfnamefont {S.}~\bibnamefont
  {Broadbent}}\ and\ \bibinfo {author} {\bibnamefont {Hammersley}},\ }\bibfield
   {title} {\bibinfo {title} {Percolation processes i. crystals and mazes},\
  }\href {https://doi.org/https://doi.org/10.1017/S0305004100032680} {\bibfield
   {journal} {\bibinfo  {journal} {Mathematical Proceedings of the Cambridge
  Philosophical Society}\ }\textbf {\bibinfo {volume} {53}},\ \bibinfo {pages}
  {629} (\bibinfo {year} {1957})}\BibitemShut {NoStop}%
\bibitem [{\citenamefont {Hunt}\ \emph {et~al.}(2014)\citenamefont {Hunt},
  \citenamefont {Ewing},\ and\ \citenamefont {Ghanbarian}}]{Hunt}%
  \BibitemOpen
  \bibfield  {author} {\bibinfo {author} {\bibfnamefont {A.}~\bibnamefont
  {Hunt}}, \bibinfo {author} {\bibfnamefont {R.}~\bibnamefont {Ewing}},\ and\
  \bibinfo {author} {\bibfnamefont {B.}~\bibnamefont {Ghanbarian}},\
  }\href@noop {} {\emph {\bibinfo {title} {Percolation Theory for Flow in
  Porous Media}}},\ \bibinfo {edition} {3rd}\ ed.\ (\bibinfo  {publisher}
  {Springer Cham},\ \bibinfo {year} {2014})\BibitemShut {NoStop}%
\bibitem [{\citenamefont {Albano}(1995)}]{Albano1}%
  \BibitemOpen
  \bibfield  {author} {\bibinfo {author} {\bibfnamefont {E.~V.}\ \bibnamefont
  {Albano}},\ }\bibfield  {title} {\bibinfo {title} {Spreading analysis and
  finite-size scaling study of the critical behavior of a forest fire model
  with immune trees},\ }\href
  {https://doi.org/https://doi.org/10.1016/0378-4371(95)00015-Y} {\bibfield
  {journal} {\bibinfo  {journal} {Physica A: Statistical Mechanics and its
  Applications}\ }\textbf {\bibinfo {volume} {216}},\ \bibinfo {pages} {213}
  (\bibinfo {year} {1995})}\BibitemShut {NoStop}%
\bibitem [{\citenamefont {Guisoni}\ \emph {et~al.}(2011)\citenamefont
  {Guisoni}, \citenamefont {Loscar},\ and\ \citenamefont {Albano}}]{Guisoni}%
  \BibitemOpen
  \bibfield  {author} {\bibinfo {author} {\bibfnamefont {N.}~\bibnamefont
  {Guisoni}}, \bibinfo {author} {\bibfnamefont {E.~S.}\ \bibnamefont
  {Loscar}},\ and\ \bibinfo {author} {\bibfnamefont {E.~V.}\ \bibnamefont
  {Albano}},\ }\bibfield  {title} {\bibinfo {title} {Phase diagram and critical
  behavior of a forest-fire model in a gradient of immunity},\ }\href
  {https://doi.org/10.1103/PhysRevE.83.011125} {\bibfield  {journal} {\bibinfo
  {journal} {Phys. Rev. E}\ }\textbf {\bibinfo {volume} {83}},\ \bibinfo
  {pages} {011125} (\bibinfo {year} {2011})}\BibitemShut {NoStop}%
\bibitem [{\citenamefont {Moore}\ and\ \citenamefont {Newman}(2000)}]{Moore}%
  \BibitemOpen
  \bibfield  {author} {\bibinfo {author} {\bibfnamefont {C.}~\bibnamefont
  {Moore}}\ and\ \bibinfo {author} {\bibfnamefont {M.}~\bibnamefont {Newman}},\
  }\bibfield  {title} {\bibinfo {title} {Epidemics and percolation in
  small-world networks},\ }\href {https://doi.org/10.1103/PhysRevE.61.5678}
  {\bibfield  {journal} {\bibinfo  {journal} {Physical review. E, Statistical
  physics, plasmas, fluids, and related interdisciplinary topics}\ }\textbf
  {\bibinfo {volume} {61}},\ \bibinfo {pages} {5678} (\bibinfo {year}
  {2000})}\BibitemShut {NoStop}%
\bibitem [{\citenamefont {Ziff}(2021)}]{Ziff2}%
  \BibitemOpen
  \bibfield  {author} {\bibinfo {author} {\bibfnamefont {R.~M.}\ \bibnamefont
  {Ziff}},\ }\bibfield  {title} {\bibinfo {title} {Percolation and the
  pandemic},\ }\href
  {https://doi.org/https://doi.org/10.1016/j.physa.2020.125723} {\bibfield
  {journal} {\bibinfo  {journal} {Physica A: Statistical Mechanics and its
  Applications}\ }\textbf {\bibinfo {volume} {568}},\ \bibinfo {pages} {125723}
  (\bibinfo {year} {2021})}\BibitemShut {NoStop}%
\bibitem [{\citenamefont {Miller}(2009)}]{Miller}%
  \BibitemOpen
  \bibfield  {author} {\bibinfo {author} {\bibfnamefont {J.~C.}\ \bibnamefont
  {Miller}},\ }\bibfield  {title} {\bibinfo {title} {Percolation and epidemics
  in random clustered networks},\ }\href
  {https://doi.org/10.1103/PhysRevE.80.020901} {\bibfield  {journal} {\bibinfo
  {journal} {Phys. Rev. E}\ }\textbf {\bibinfo {volume} {80}},\ \bibinfo
  {pages} {020901(R)} (\bibinfo {year} {2009})}\BibitemShut {NoStop}%
\bibitem [{\citenamefont {Saberi}(2015)}]{Saberi1}%
  \BibitemOpen
  \bibfield  {author} {\bibinfo {author} {\bibfnamefont {A.~A.}\ \bibnamefont
  {Saberi}},\ }\bibfield  {title} {\bibinfo {title} {Recent advances in
  percolation theory and its applications},\ }\href
  {https://doi.org/https://doi.org/10.1016/j.physrep.2015.03.003} {\bibfield
  {journal} {\bibinfo  {journal} {Physics Reports}\ }\textbf {\bibinfo {volume}
  {578}},\ \bibinfo {pages} {1} (\bibinfo {year} {2015})}\BibitemShut {NoStop}%
\bibitem [{\citenamefont {Stauffer}\ and\ \citenamefont
  {Aharony}(1992)}]{Stauffer}%
  \BibitemOpen
  \bibfield  {author} {\bibinfo {author} {\bibfnamefont {D.}~\bibnamefont
  {Stauffer}}\ and\ \bibinfo {author} {\bibfnamefont {A.}~\bibnamefont
  {Aharony}},\ }\href@noop {} {\emph {\bibinfo {title} {Introduction To
  Percolation Theory}}},\ \bibinfo {edition} {2nd}\ ed.\ (\bibinfo  {publisher}
  {Taylor {\&} Francis},\ \bibinfo {address} {London},\ \bibinfo {year}
  {1992})\BibitemShut {NoStop}%
\bibitem [{\citenamefont {Ali~Saberi}(2013)}]{Saberi3}%
  \BibitemOpen
  \bibfield  {author} {\bibinfo {author} {\bibfnamefont {A.}~\bibnamefont
  {Ali~Saberi}},\ }\bibfield  {title} {\bibinfo {title} {Percolation
  description of the global topography of earth and the moon},\ }\href
  {https://doi.org/10.1103/PhysRevLett.110.178501} {\bibfield  {journal}
  {\bibinfo  {journal} {Phys. Rev. Lett.}\ }\textbf {\bibinfo {volume} {110}},\
  \bibinfo {pages} {178501} (\bibinfo {year} {2013})}\BibitemShut {NoStop}%
\bibitem [{\citenamefont {Ball}\ \emph {et~al.}(1994)\citenamefont {Ball},
  \citenamefont {Phillips}, \citenamefont {Callahan},\ and\ \citenamefont
  {Sauerbrey}}]{Ball}%
  \BibitemOpen
  \bibfield  {author} {\bibinfo {author} {\bibfnamefont {Z.}~\bibnamefont
  {Ball}}, \bibinfo {author} {\bibfnamefont {H.~M.}\ \bibnamefont {Phillips}},
  \bibinfo {author} {\bibfnamefont {D.~L.}\ \bibnamefont {Callahan}},\ and\
  \bibinfo {author} {\bibfnamefont {R.}~\bibnamefont {Sauerbrey}},\ }\bibfield
  {title} {\bibinfo {title} {Percolative metal-insulator transition in excimer
  laser irradiated polyimide},\ }\href
  {https://doi.org/10.1103/PhysRevLett.73.2099} {\bibfield  {journal} {\bibinfo
   {journal} {Phys. Rev. Lett.}\ }\textbf {\bibinfo {volume} {73}},\ \bibinfo
  {pages} {2099} (\bibinfo {year} {1994})}\BibitemShut {NoStop}%
\bibitem [{\citenamefont {Dotsenko}\ \emph {et~al.}(1993)\citenamefont
  {Dotsenko}, \citenamefont {Windey}, \citenamefont {Harris}, \citenamefont
  {Marinari}, \citenamefont {Martinec},\ and\ \citenamefont
  {Picco}}]{Dotsenko}%
  \BibitemOpen
  \bibfield  {author} {\bibinfo {author} {\bibfnamefont {V.~S.}\ \bibnamefont
  {Dotsenko}}, \bibinfo {author} {\bibfnamefont {P.}~\bibnamefont {Windey}},
  \bibinfo {author} {\bibfnamefont {G.}~\bibnamefont {Harris}}, \bibinfo
  {author} {\bibfnamefont {E.}~\bibnamefont {Marinari}}, \bibinfo {author}
  {\bibfnamefont {E.}~\bibnamefont {Martinec}},\ and\ \bibinfo {author}
  {\bibfnamefont {M.}~\bibnamefont {Picco}},\ }\bibfield  {title} {\bibinfo
  {title} {Critical and topological properties of cluster boundaries in the 3d
  ising model},\ }\href {https://doi.org/10.1103/PhysRevLett.71.811} {\bibfield
   {journal} {\bibinfo  {journal} {Phys. Rev. Lett.}\ }\textbf {\bibinfo
  {volume} {71}},\ \bibinfo {pages} {811} (\bibinfo {year} {1993})}\BibitemShut
  {NoStop}%
\bibitem [{\citenamefont {Wang}\ \emph {et~al.}(2013)\citenamefont {Wang},
  \citenamefont {Zhou}, \citenamefont {Zhang}, \citenamefont {Garoni},\ and\
  \citenamefont {Deng}}]{Wang}%
  \BibitemOpen
  \bibfield  {author} {\bibinfo {author} {\bibfnamefont {J.}~\bibnamefont
  {Wang}}, \bibinfo {author} {\bibfnamefont {Z.}~\bibnamefont {Zhou}}, \bibinfo
  {author} {\bibfnamefont {W.}~\bibnamefont {Zhang}}, \bibinfo {author}
  {\bibfnamefont {T.~M.}\ \bibnamefont {Garoni}},\ and\ \bibinfo {author}
  {\bibfnamefont {Y.}~\bibnamefont {Deng}},\ }\bibfield  {title} {\bibinfo
  {title} {Bond and site percolation in three dimensions},\ }\href
  {https://doi.org/10.1103/PhysRevE.87.052107} {\bibfield  {journal} {\bibinfo
  {journal} {Phys. Rev. E}\ }\textbf {\bibinfo {volume} {87}},\ \bibinfo
  {pages} {052107} (\bibinfo {year} {2013})}\BibitemShut {NoStop}%
\bibitem [{\citenamefont {Manna}\ and\ \citenamefont {Ziff}(2020)}]{Manna}%
  \BibitemOpen
  \bibfield  {author} {\bibinfo {author} {\bibfnamefont {S.~S.}\ \bibnamefont
  {Manna}}\ and\ \bibinfo {author} {\bibfnamefont {R.~M.}\ \bibnamefont
  {Ziff}},\ }\bibfield  {title} {\bibinfo {title} {Bond percolation between $k$
  separated points on a square lattice},\ }\href
  {https://doi.org/10.1103/PhysRevE.101.062143} {\bibfield  {journal} {\bibinfo
   {journal} {Phys. Rev. E}\ }\textbf {\bibinfo {volume} {101}},\ \bibinfo
  {pages} {062143} (\bibinfo {year} {2020})}\BibitemShut {NoStop}%
\bibitem [{\citenamefont {Tarasevich}\ and\ \citenamefont {Van~der
  Marck}(1999)}]{Tarasevich}%
  \BibitemOpen
  \bibfield  {author} {\bibinfo {author} {\bibfnamefont {Y.~Y.}\ \bibnamefont
  {Tarasevich}}\ and\ \bibinfo {author} {\bibfnamefont {S.~C.}\ \bibnamefont
  {Van~der Marck}},\ }\bibfield  {title} {\bibinfo {title} {An investigation of
  site-bond percolation on many lattices},\ }\href
  {https://doi.org/10.1142/s0129183199000978} {\bibfield  {journal} {\bibinfo
  {journal} {International Journal of Modern Physics C}\ }\textbf {\bibinfo
  {volume} {10}},\ \bibinfo {pages} {1193} (\bibinfo {year}
  {1999})}\BibitemShut {NoStop}%
\bibitem [{\citenamefont {Takeuchi}\ \emph {et~al.}(2007)\citenamefont
  {Takeuchi}, \citenamefont {Kuroda}, \citenamefont {Chat\'e},\ and\
  \citenamefont {Sano}}]{Takeuchi}%
  \BibitemOpen
  \bibfield  {author} {\bibinfo {author} {\bibfnamefont {K.~A.}\ \bibnamefont
  {Takeuchi}}, \bibinfo {author} {\bibfnamefont {M.}~\bibnamefont {Kuroda}},
  \bibinfo {author} {\bibfnamefont {H.}~\bibnamefont {Chat\'e}},\ and\ \bibinfo
  {author} {\bibfnamefont {M.}~\bibnamefont {Sano}},\ }\bibfield  {title}
  {\bibinfo {title} {Directed percolation criticality in turbulent liquid
  crystals},\ }\href
  {https://doi.org/https://doi.org/10.1103/PhysRevLett.99.234503} {\bibfield
  {journal} {\bibinfo  {journal} {Phys. Rev. Lett.}\ }\textbf {\bibinfo
  {volume} {99}},\ \bibinfo {pages} {234503} (\bibinfo {year}
  {2007})}\BibitemShut {NoStop}%
\bibitem [{\citenamefont {Adler}(1991)}]{Adler}%
  \BibitemOpen
  \bibfield  {author} {\bibinfo {author} {\bibfnamefont {J.}~\bibnamefont
  {Adler}},\ }\bibfield  {title} {\bibinfo {title} {Bootstrap percolation},\
  }\href {https://doi.org/https://doi.org/10.1016/0378-4371(91)90295-N}
  {\bibfield  {journal} {\bibinfo  {journal} {Physica A: Statistical Mechanics
  and its Applications}\ }\textbf {\bibinfo {volume} {171}},\ \bibinfo {pages}
  {453} (\bibinfo {year} {1991})}\BibitemShut {NoStop}%
\bibitem [{\citenamefont {Achlioptas}\ \emph {et~al.}(2009)\citenamefont
  {Achlioptas}, \citenamefont {D'Souza},\ and\ \citenamefont
  {Spencer}}]{Achlioptas}%
  \BibitemOpen
  \bibfield  {author} {\bibinfo {author} {\bibfnamefont {D.}~\bibnamefont
  {Achlioptas}}, \bibinfo {author} {\bibfnamefont {R.~M.}\ \bibnamefont
  {D'Souza}},\ and\ \bibinfo {author} {\bibfnamefont {J.}~\bibnamefont
  {Spencer}},\ }\bibfield  {title} {\bibinfo {title} {Explosive percolation in
  random networks},\ }\href
  {https://doi.org/https://doi.org/10.1126/science.1167782} {\bibfield
  {journal} {\bibinfo  {journal} {Science}\ }\textbf {\bibinfo {volume}
  {323}},\ \bibinfo {pages} {1453} (\bibinfo {year} {2009})}\BibitemShut
  {NoStop}%
\bibitem [{\citenamefont {Coniglio}\ \emph {et~al.}(1979)\citenamefont
  {Coniglio}, \citenamefont {Stanley},\ and\ \citenamefont {Klein}}]{Coniglio}%
  \BibitemOpen
  \bibfield  {author} {\bibinfo {author} {\bibfnamefont {A.}~\bibnamefont
  {Coniglio}}, \bibinfo {author} {\bibfnamefont {H.~E.}\ \bibnamefont
  {Stanley}},\ and\ \bibinfo {author} {\bibfnamefont {W.}~\bibnamefont
  {Klein}},\ }\bibfield  {title} {\bibinfo {title} {Site-bond
  correlated-percolation problem: A statistical mechanical model of polymer
  gelation},\ }\href {https://doi.org/10.1103/PhysRevLett.42.518} {\bibfield
  {journal} {\bibinfo  {journal} {Phys. Rev. Lett.}\ }\textbf {\bibinfo
  {volume} {42}},\ \bibinfo {pages} {518} (\bibinfo {year} {1979})}\BibitemShut
  {NoStop}%
\bibitem [{\citenamefont {Hall}(1985)}]{Hall1985}%
  \BibitemOpen
  \bibfield  {author} {\bibinfo {author} {\bibfnamefont {P.}~\bibnamefont
  {Hall}},\ }\bibfield  {title} {\bibinfo {title} {{On Continuum
  Percolation}},\ }\href {https://doi.org/10.1214/aop/1176992809} {\bibfield
  {journal} {\bibinfo  {journal} {The Annals of Probability}\ }\textbf
  {\bibinfo {volume} {13}},\ \bibinfo {pages} {1250 } (\bibinfo {year}
  {1985})}\BibitemShut {NoStop}%
\bibitem [{\citenamefont {Mertens}\ and\ \citenamefont
  {Moore}(2012)}]{Mertens}%
  \BibitemOpen
  \bibfield  {author} {\bibinfo {author} {\bibfnamefont {S.}~\bibnamefont
  {Mertens}}\ and\ \bibinfo {author} {\bibfnamefont {C.}~\bibnamefont
  {Moore}},\ }\bibfield  {title} {\bibinfo {title} {Continuum percolation
  thresholds in two dimensions},\ }\href
  {https://doi.org/10.1103/PhysRevE.86.061109} {\bibfield  {journal} {\bibinfo
  {journal} {Phys. Rev. E}\ }\textbf {\bibinfo {volume} {86}},\ \bibinfo
  {pages} {061109} (\bibinfo {year} {2012})}\BibitemShut {NoStop}%
\bibitem [{\citenamefont {Deng}\ and\ \citenamefont {Bl\"ote}(2005)}]{Deng}%
  \BibitemOpen
  \bibfield  {author} {\bibinfo {author} {\bibfnamefont {Y.}~\bibnamefont
  {Deng}}\ and\ \bibinfo {author} {\bibfnamefont {H.~W.~J.}\ \bibnamefont
  {Bl\"ote}},\ }\bibfield  {title} {\bibinfo {title} {Monte carlo study of the
  site-percolation model in two and three dimensions},\ }\href
  {https://doi.org/10.1103/PhysRevE.72.016126} {\bibfield  {journal} {\bibinfo
  {journal} {Phys. Rev. E}\ }\textbf {\bibinfo {volume} {72}},\ \bibinfo
  {pages} {016126} (\bibinfo {year} {2005})}\BibitemShut {NoStop}%
\bibitem [{\citenamefont {Ballesteros}\ \emph {et~al.}(1999)\citenamefont
  {Ballesteros}, \citenamefont {Fern\'{a}ndez}, \citenamefont {Mart\'{i}n-Mayor},
  \citenamefont {Sudupe}, \citenamefont {Parisi},\ and\ \citenamefont
  {Ruiz-Lorenzo}}]{Ballesteros}%
  \BibitemOpen
  \bibfield  {author} {\bibinfo {author} {\bibfnamefont {H.~G.}\ \bibnamefont
  {Ballesteros}}, \bibinfo {author} {\bibfnamefont {L.~A.}\ \bibnamefont
  {Fern\'{a}ndez}}, \bibinfo {author} {\bibfnamefont {V.}~\bibnamefont
  {Mart\'{i}n-Mayor}}, \bibinfo {author} {\bibfnamefont {A.~M.}\ \bibnamefont
  {Sudupe}}, \bibinfo {author} {\bibfnamefont {G.}~\bibnamefont {Parisi}},\
  and\ \bibinfo {author} {\bibfnamefont {J.~J.}\ \bibnamefont {Ruiz-Lorenzo}},\
  }\bibfield  {title} {\bibinfo {title} {Scaling corrections: site percolation
  and ising model in three dimensions},\ }\href
  {https://doi.org/10.1088/0305-4470/32/1/004} {\bibfield  {journal} {\bibinfo
  {journal} {Journal of Physics A: Mathematical and General}\ }\textbf
  {\bibinfo {volume} {32}},\ \bibinfo {pages} {1} (\bibinfo {year}
  {1999})}\BibitemShut {NoStop}%
\bibitem [{\citenamefont {Gonz{\'{a}}lez}\ \emph {et~al.}(2016)\citenamefont
  {Gonz{\'{a}}lez}, \citenamefont {Centres}, \citenamefont {Lebrecht},\ and\
  \citenamefont {Ramirez-Pastor}}]{Gonzalez}%
  \BibitemOpen
  \bibfield  {author} {\bibinfo {author} {\bibfnamefont {M.~I.}\ \bibnamefont
  {Gonz{\'{a}}lez}}, \bibinfo {author} {\bibfnamefont {P.~M.}\ \bibnamefont
  {Centres}}, \bibinfo {author} {\bibfnamefont {W.}~\bibnamefont {Lebrecht}},\
  and\ \bibinfo {author} {\bibfnamefont {A.~J.}\ \bibnamefont
  {Ramirez-Pastor}},\ }\bibfield  {title} {\bibinfo {title} {Site-bond
  percolation on simple cubic lattices: numerical simulation and analytical
  approach},\ }\href {https://doi.org/10.1088/1742-5468/2016/09/093210}
  {\bibfield  {journal} {\bibinfo  {journal} {Journal of Statistical Mechanics:
  Theory and Experiment}\ }\textbf {\bibinfo {volume} {2016}},\ \bibinfo
  {pages} {093210} (\bibinfo {year} {2016})}\BibitemShut {NoStop}%
\bibitem [{\citenamefont {Lorenz}\ and\ \citenamefont {Ziff}(1998)}]{Lorentz}%
  \BibitemOpen
  \bibfield  {author} {\bibinfo {author} {\bibfnamefont {C.~D.}\ \bibnamefont
  {Lorenz}}\ and\ \bibinfo {author} {\bibfnamefont {R.~M.}\ \bibnamefont
  {Ziff}},\ }\bibfield  {title} {\bibinfo {title} {Precise determination of the
  bond percolation thresholds and finite-size scaling corrections for the sc,
  fcc, and bcc lattices},\ }\href {https://doi.org/10.1103/PhysRevE.57.230}
  {\bibfield  {journal} {\bibinfo  {journal} {Phys. Rev. E}\ }\textbf {\bibinfo
  {volume} {57}},\ \bibinfo {pages} {230} (\bibinfo {year} {1998})}\BibitemShut
  {NoStop}%
\bibitem [{\citenamefont {Sommers}\ \emph {et~al.}(2023)\citenamefont
  {Sommers}, \citenamefont {Gullans},\ and\ \citenamefont {Huse}}]{Sommers}%
  \BibitemOpen
  \bibfield  {author} {\bibinfo {author} {\bibfnamefont {G.~M.}\ \bibnamefont
  {Sommers}}, \bibinfo {author} {\bibfnamefont {M.~J.}\ \bibnamefont
  {Gullans}},\ and\ \bibinfo {author} {\bibfnamefont {D.~A.}\ \bibnamefont
  {Huse}},\ }\bibfield  {title} {\bibinfo {title} {Self-dual quasiperiodic
  percolation},\ }\href {https://doi.org/10.1103/PhysRevE.107.024137}
  {\bibfield  {journal} {\bibinfo  {journal} {Phys. Rev. E}\ }\textbf {\bibinfo
  {volume} {107}},\ \bibinfo {pages} {024137} (\bibinfo {year}
  {2023})}\BibitemShut {NoStop}%
\bibitem [{\citenamefont {Liu}\ and\ \citenamefont
  {Regenauer-Lieb}(2011)}]{Liu}%
  \BibitemOpen
  \bibfield  {author} {\bibinfo {author} {\bibfnamefont {J.}~\bibnamefont
  {Liu}}\ and\ \bibinfo {author} {\bibfnamefont {K.}~\bibnamefont
  {Regenauer-Lieb}},\ }\bibfield  {title} {\bibinfo {title} {Application of
  percolation theory to microtomography of structured media: Percolation
  threshold, critical exponents, and upscaling},\ }\href
  {https://doi.org/10.1103/PhysRevE.83.016106} {\bibfield  {journal} {\bibinfo
  {journal} {Phys. Rev. E}\ }\textbf {\bibinfo {volume} {83}},\ \bibinfo
  {pages} {016106} (\bibinfo {year} {2011})}\BibitemShut {NoStop}%
\bibitem [{\citenamefont {Kirkpatrick}(1976)}]{Kirkpatrick}%
  \BibitemOpen
  \bibfield  {author} {\bibinfo {author} {\bibfnamefont {S.}~\bibnamefont
  {Kirkpatrick}},\ }\bibfield  {title} {\bibinfo {title} {Percolation phenomena
  in higher dimensions: Approach to the mean-field limit},\ }\href
  {https://doi.org/10.1103/PhysRevLett.36.69} {\bibfield  {journal} {\bibinfo
  {journal} {Phys. Rev. Lett.}\ }\textbf {\bibinfo {volume} {36}},\ \bibinfo
  {pages} {69} (\bibinfo {year} {1976})}\BibitemShut {NoStop}%
\bibitem [{\citenamefont {Kundu}\ and\ \citenamefont {Mandal}(2021)}]{Kundu2}%
  \BibitemOpen
  \bibfield  {author} {\bibinfo {author} {\bibfnamefont {S.}~\bibnamefont
  {Kundu}}\ and\ \bibinfo {author} {\bibfnamefont {D.}~\bibnamefont {Mandal}},\
  }\bibfield  {title} {\bibinfo {title} {Breaking universality in random
  sequential adsorption on a square lattice with long-range correlated
  defects},\ }\href {https://doi.org/10.1103/PhysRevE.87.052107} {\bibfield
  {journal} {\bibinfo  {journal} {Phys. Rev. E}\ }\textbf {\bibinfo {volume}
  {103}},\ \bibinfo {pages} {042134} (\bibinfo {year} {2021})}\BibitemShut
  {NoStop}%
\bibitem [{\citenamefont {Dalton}\ \emph {et~al.}(1964)\citenamefont {Dalton},
  \citenamefont {Domb},\ and\ \citenamefont {Sykes}}]{Dalton1964}%
  \BibitemOpen
  \bibfield  {author} {\bibinfo {author} {\bibfnamefont {N.~W.}\ \bibnamefont
  {Dalton}}, \bibinfo {author} {\bibfnamefont {C.}~\bibnamefont {Domb}},\ and\
  \bibinfo {author} {\bibfnamefont {M.~F.}\ \bibnamefont {Sykes}},\ }\bibfield
  {title} {\bibinfo {title} {Dependence of critical concentration of a dilute
  ferromagnet on the range of interaction},\ }\href
  {https://doi.org/10.1088/0370-1328/83/3/118} {\bibfield  {journal} {\bibinfo
  {journal} {Proceedings of the Physical Society}\ }\textbf {\bibinfo {volume}
  {83}},\ \bibinfo {pages} {496} (\bibinfo {year} {1964})}\BibitemShut
  {NoStop}%
\bibitem [{\citenamefont {Domb}\ and\ \citenamefont
  {Dalton}(1966{\natexlab{a}})}]{Domb1966}%
  \BibitemOpen
  \bibfield  {author} {\bibinfo {author} {\bibfnamefont {C.}~\bibnamefont
  {Domb}}\ and\ \bibinfo {author} {\bibfnamefont {N.~W.}\ \bibnamefont
  {Dalton}},\ }\bibfield  {title} {\bibinfo {title} {Crystal statistics with
  long-range forces: I. the equivalent neighbour model},\ }\href
  {https://doi.org/10.1088/0370-1328/89/4/311} {\bibfield  {journal} {\bibinfo
  {journal} {Proceedings of the Physical Society}\ }\textbf {\bibinfo {volume}
  {89}},\ \bibinfo {pages} {859} (\bibinfo {year}
  {1966}{\natexlab{a}})}\BibitemShut {NoStop}%
\bibitem [{\citenamefont {Gouker}\ and\ \citenamefont
  {Family}(1983)}]{Gouker1983}%
  \BibitemOpen
  \bibfield  {author} {\bibinfo {author} {\bibfnamefont {M.}~\bibnamefont
  {Gouker}}\ and\ \bibinfo {author} {\bibfnamefont {F.}~\bibnamefont
  {Family}},\ }\bibfield  {title} {\bibinfo {title} {Evidence for classical
  critical behavior in long-range site percolation},\ }\href
  {https://doi.org/10.1103/PhysRevB.28.1449} {\bibfield  {journal} {\bibinfo
  {journal} {Phys. Rev. B}\ }\textbf {\bibinfo {volume} {28}},\ \bibinfo
  {pages} {1449} (\bibinfo {year} {1983})}\BibitemShut {NoStop}%
\bibitem [{\citenamefont {d'Iribarne}\ \emph {et~al.}(1999)\citenamefont
  {d'Iribarne}, \citenamefont {Rasigni},\ and\ \citenamefont
  {Rasigni}}]{Diribarne1999}%
  \BibitemOpen
  \bibfield  {author} {\bibinfo {author} {\bibfnamefont {C.}~\bibnamefont
  {d'Iribarne}}, \bibinfo {author} {\bibfnamefont {M.}~\bibnamefont
  {Rasigni}},\ and\ \bibinfo {author} {\bibfnamefont {G.}~\bibnamefont
  {Rasigni}},\ }\bibfield  {title} {\bibinfo {title} {From lattice long-range
  percolation to the continuum one},\ }\href
  {https://doi.org/https://doi.org/10.1016/S0375-9601(99)00585-X} {\bibfield
  {journal} {\bibinfo  {journal} {Physics Letters A}\ }\textbf {\bibinfo
  {volume} {263}},\ \bibinfo {pages} {65} (\bibinfo {year} {1999})}\BibitemShut
  {NoStop}%
\bibitem [{\citenamefont {Malarz}\ and\ \citenamefont
  {Galam}(2005)}]{Malarz2005}%
  \BibitemOpen
  \bibfield  {author} {\bibinfo {author} {\bibfnamefont {K.}~\bibnamefont
  {Malarz}}\ and\ \bibinfo {author} {\bibfnamefont {S.}~\bibnamefont {Galam}},\
  }\bibfield  {title} {\bibinfo {title} {Square-lattice site percolation at
  increasing ranges of neighbor bonds},\ }\href
  {https://doi.org/10.1103/PhysRevE.71.016125} {\bibfield  {journal} {\bibinfo
  {journal} {Phys. Rev. E}\ }\textbf {\bibinfo {volume} {71}},\ \bibinfo
  {pages} {016125} (\bibinfo {year} {2005})}\BibitemShut {NoStop}%
\bibitem [{\citenamefont {Malarz}(2021)}]{Malarz2021}%
  \BibitemOpen
  \bibfield  {author} {\bibinfo {author} {\bibfnamefont {K.}~\bibnamefont
  {Malarz}},\ }\bibfield  {title} {\bibinfo {title} {Percolation thresholds on
  a triangular lattice for neighborhoods containing sites up to the fifth
  coordination zone},\ }\bibfield  {journal} {\bibinfo  {journal} {Physical
  Review E}\ }\textbf {\bibinfo {volume} {103}},\ \bibinfo
  {pages} {052107} (\bibinfo {year} {2021})\BibitemShut {NoStop}%
\bibitem [{\citenamefont {Malarz}(2022)}]{Malarz2022}%
  \BibitemOpen
  \bibfield  {author} {\bibinfo {author} {\bibfnamefont {K.}~\bibnamefont
  {Malarz}},\ }\bibfield  {title} {\bibinfo {title} {Random site percolation on
  honeycomb lattices with complex neighborhoods},\ }\href
  {https://doi.org/10.1063/5.0099066} {\bibfield  {journal} {\bibinfo
  {journal} {Chaos: An Interdisciplinary Journal of Nonlinear Science}\
  }\textbf {\bibinfo {volume} {32}},\ \bibinfo {pages} {083123} (\bibinfo
  {year} {2022})}\BibitemShut {NoStop}%
\bibitem [{\citenamefont {Xun}\ \emph {et~al.}(2021)\citenamefont {Xun},
  \citenamefont {Hao},\ and\ \citenamefont {Ziff}}]{Xun1}%
  \BibitemOpen
  \bibfield  {author} {\bibinfo {author} {\bibfnamefont {Z.}~\bibnamefont
  {Xun}}, \bibinfo {author} {\bibfnamefont {D.}~\bibnamefont {Hao}},\ and\
  \bibinfo {author} {\bibfnamefont {R.~M.}\ \bibnamefont {Ziff}},\ }\bibfield
  {title} {\bibinfo {title} {Site percolation on square and simple cubic
  lattices with extended neighborhoods and their continuum limit},\ }\href
  {https://doi.org/10.1103/PhysRevE.103.022126} {\bibfield  {journal} {\bibinfo
   {journal} {Phys. Rev. E}\ }\textbf {\bibinfo {volume} {103}},\ \bibinfo
  {pages} {022126} (\bibinfo {year} {2021})}\BibitemShut {NoStop}%
\bibitem [{\citenamefont {Xun}\ and\ \citenamefont {Ziff}(2020)}]{Xun2}%
  \BibitemOpen
  \bibfield  {author} {\bibinfo {author} {\bibfnamefont {Z.}~\bibnamefont
  {Xun}}\ and\ \bibinfo {author} {\bibfnamefont {R.~M.}\ \bibnamefont {Ziff}},\
  }\bibfield  {title} {\bibinfo {title} {Bond percolation on simple cubic
  lattices with extended neighborhoods},\ }\href
  {https://doi.org/10.1103/PhysRevE.102.012102} {\bibfield  {journal} {\bibinfo
   {journal} {Phys. Rev. E}\ }\textbf {\bibinfo {volume} {102}},\ \bibinfo
  {pages} {012102} (\bibinfo {year} {2020})}\BibitemShut {NoStop}%
\bibitem [{\citenamefont {Xun}\ \emph {et~al.}(2022)\citenamefont {Xun},
  \citenamefont {Hao},\ and\ \citenamefont {Ziff}}]{Xun3}%
  \BibitemOpen
  \bibfield  {author} {\bibinfo {author} {\bibfnamefont {Z.}~\bibnamefont
  {Xun}}, \bibinfo {author} {\bibfnamefont {D.}~\bibnamefont {Hao}},\ and\
  \bibinfo {author} {\bibfnamefont {R.~M.}\ \bibnamefont {Ziff}},\ }\bibfield
  {title} {\bibinfo {title} {Site and bond percolation thresholds on regular
  lattices with compact extended-range neighborhoods in two and three
  dimensions},\ }\href {https://doi.org/10.1103/PhysRevE.105.024105} {\bibfield
   {journal} {\bibinfo  {journal} {Phys. Rev. E}\ }\textbf {\bibinfo {volume}
  {105}},\ \bibinfo {pages} {024105} (\bibinfo {year} {2022})}\BibitemShut
  {NoStop}%
\bibitem [{\citenamefont {Mitra}\ \emph {et~al.}(2019)\citenamefont {Mitra},
  \citenamefont {Saha},\ and\ \citenamefont {Sensharma}}]{Sayantan1}%
  \BibitemOpen
  \bibfield  {author} {\bibinfo {author} {\bibfnamefont {S.}~\bibnamefont
  {Mitra}}, \bibinfo {author} {\bibfnamefont {D.}~\bibnamefont {Saha}},\ and\
  \bibinfo {author} {\bibfnamefont {A.}~\bibnamefont {Sensharma}},\ }\bibfield
  {title} {\bibinfo {title} {Percolation in a distorted square lattice},\
  }\href {https://doi.org/10.1103/PhysRevE.99.012117} {\bibfield  {journal}
  {\bibinfo  {journal} {Phys. Rev. E}\ }\textbf {\bibinfo {volume} {99}},\
  \bibinfo {pages} {012117} (\bibinfo {year} {2019})}\BibitemShut {NoStop}%
\bibitem [{\citenamefont {Mitra}\ \emph {et~al.}(2022)\citenamefont {Mitra},
  \citenamefont {Saha},\ and\ \citenamefont {Sensharma}}]{Sayantan2}%
  \BibitemOpen
  \bibfield  {author} {\bibinfo {author} {\bibfnamefont {S.}~\bibnamefont
  {Mitra}}, \bibinfo {author} {\bibfnamefont {D.}~\bibnamefont {Saha}},\ and\
  \bibinfo {author} {\bibfnamefont {A.}~\bibnamefont {Sensharma}},\ }\bibfield
  {title} {\bibinfo {title} {Percolation in a simple cubic lattice with
  distortion},\ }\href {https://doi.org/10.1103/PhysRevE.106.034109} {\bibfield
   {journal} {\bibinfo  {journal} {Phys. Rev. E}\ }\textbf {\bibinfo {volume}
  {106}},\ \bibinfo {pages} {034109} (\bibinfo {year} {2022})}\BibitemShut
  {NoStop}%
\bibitem [{\citenamefont {Jang}\ and\ \citenamefont {Yu}(2019)}]{Jang}%
  \BibitemOpen
  \bibfield  {author} {\bibinfo {author} {\bibfnamefont {H.}~\bibnamefont
  {Jang}}\ and\ \bibinfo {author} {\bibfnamefont {U.}~\bibnamefont {Yu}},\
  }\bibfield  {title} {\bibinfo {title} {Universality class of the percolation
  in two-dimensional lattices with distortion},\ }\href
  {https://doi.org/https://doi.org/10.1016/j.physa.2019.121139} {\bibfield
  {journal} {\bibinfo  {journal} {Physica A: Statistical Mechanics and its
  Applications}\ }\textbf {\bibinfo {volume} {527}},\ \bibinfo {pages} {121139}
  (\bibinfo {year} {2019})}\BibitemShut {NoStop}%
\bibitem [{\citenamefont {Newman}\ and\ \citenamefont {Ziff}(2001)}]{Newman2}%
  \BibitemOpen
  \bibfield  {author} {\bibinfo {author} {\bibfnamefont {M.~E.~J.}\
  \bibnamefont {Newman}}\ and\ \bibinfo {author} {\bibfnamefont {R.~M.}\
  \bibnamefont {Ziff}},\ }\bibfield  {title} {\bibinfo {title} {Fast monte
  carlo algorithm for site or bond percolation},\ }\href
  {https://doi.org/10.1103/PhysRevE.64.016706} {\bibfield  {journal} {\bibinfo
  {journal} {Phys. Rev. E}\ }\textbf {\bibinfo {volume} {64}},\ \bibinfo
  {pages} {016706} (\bibinfo {year} {2001})}\BibitemShut {NoStop}%
\bibitem [{\citenamefont {Newman}\ and\ \citenamefont {Ziff}(2000)}]{Newman1}%
  \BibitemOpen
  \bibfield  {author} {\bibinfo {author} {\bibfnamefont {M.~E.~J.}\
  \bibnamefont {Newman}}\ and\ \bibinfo {author} {\bibfnamefont {R.~M.}\
  \bibnamefont {Ziff}},\ }\bibfield  {title} {\bibinfo {title} {Efficient monte
  carlo algorithm and high-precision results for percolation},\ }\href
  {https://doi.org/https://doi.org/10.1103/PhysRevLett.85.4104} {\bibfield
  {journal} {\bibinfo  {journal} {Phys. Rev. Lett.}\ }\textbf {\bibinfo
  {volume} {85}},\ \bibinfo {pages} {4104} (\bibinfo {year}
  {2000})}\BibitemShut {NoStop}%
\bibitem [{Note1()}]{Note1}%
  \BibitemOpen
  \bibinfo {note} {The random number generator used is RAN2()}\BibitemShut
  {NoStop}%
\bibitem [{\citenamefont {Domb}\ and\ \citenamefont
  {Dalton}(1966{\natexlab{b}})}]{Domb}%
  \BibitemOpen
  \bibfield  {author} {\bibinfo {author} {\bibfnamefont {C.}~\bibnamefont
  {Domb}}\ and\ \bibinfo {author} {\bibfnamefont {N.~W.}\ \bibnamefont
  {Dalton}},\ }\bibfield  {title} {\bibinfo {title} {Crystal statistics with
  long-range forces: I. the equivalent neighbour model},\ }\href
  {https://doi.org/10.1088/0370-1328/89/4/311} {\bibfield  {journal} {\bibinfo
  {journal} {Proceedings of the Physical Society}\ }\textbf {\bibinfo {volume}
  {89}},\ \bibinfo {pages} {859} (\bibinfo {year}
  {1966}{\natexlab{b}})}\BibitemShut {NoStop}%
\bibitem [{\citenamefont {Quintanilla}\ and\ \citenamefont
  {Ziff}(2007)}]{Quintanilla}%
  \BibitemOpen
  \bibfield  {author} {\bibinfo {author} {\bibfnamefont {J.~A.}\ \bibnamefont
  {Quintanilla}}\ and\ \bibinfo {author} {\bibfnamefont {R.~M.}\ \bibnamefont
  {Ziff}},\ }\bibfield  {title} {\bibinfo {title} {Asymmetry in the percolation
  thresholds of fully penetrable disks with two different radii},\ }\href
  {https://doi.org/10.1103/PhysRevE.76.051115} {\bibfield  {journal} {\bibinfo
  {journal} {Phys. Rev. E}\ }\textbf {\bibinfo {volume} {76}},\ \bibinfo
  {pages} {051115} (\bibinfo {year} {2007})}\BibitemShut {NoStop}%
\bibitem [{\citenamefont {Tarasevich}\ and\ \citenamefont
  {Eserkepov}(2020)}]{Tarasevich1}%
  \BibitemOpen
  \bibfield  {author} {\bibinfo {author} {\bibfnamefont {Y.~Y.}\ \bibnamefont
  {Tarasevich}}\ and\ \bibinfo {author} {\bibfnamefont {A.~V.}\ \bibnamefont
  {Eserkepov}},\ }\bibfield  {title} {\bibinfo {title} {Percolation thresholds
  for discorectangles: Numerical estimation for a range of aspect ratios},\
  }\href {https://doi.org/10.1103/PhysRevE.101.022108} {\bibfield  {journal}
  {\bibinfo  {journal} {Phys. Rev. E}\ }\textbf {\bibinfo {volume} {101}},\
  \bibinfo {pages} {022108} (\bibinfo {year} {2020})}\BibitemShut {NoStop}%
\bibitem [{\citenamefont {Lorenz}\ and\ \citenamefont {Ziff}(2001)}]{Lorenz1}%
  \BibitemOpen
  \bibfield  {author} {\bibinfo {author} {\bibfnamefont {C.~D.}\ \bibnamefont
  {Lorenz}}\ and\ \bibinfo {author} {\bibfnamefont {R.~M.}\ \bibnamefont
  {Ziff}},\ }\bibfield  {title} {\bibinfo {title} {Precise determination of the
  critical percolation threshold for the three-dimensional “swiss cheese”
  model using a growth algorithm},\ }\href {https://doi.org/10.1063/1.1338506}
  {\bibfield  {journal} {\bibinfo  {journal} {The Journal of Chemical Physics}\
  }\textbf {\bibinfo {volume} {114}},\ \bibinfo {pages} {3659} (\bibinfo {year}
  {2001})},\ \Eprint {https://arxiv.org/abs/https://doi.org/10.1063/1.1338506}
  {https://doi.org/10.1063/1.1338506} \BibitemShut {NoStop}%
\end{thebibliography}

%
\end{document}